\documentclass[runningheads]{llncs}
\usepackage{graphicx}

\usepackage{natbib}
\setcitestyle{square,numbers}

\usepackage[toc,page]{appendix}

\usepackage{nameref}

\usepackage{amsmath}
\usepackage{amssymb}

\usepackage{xcolor}
\usepackage{color,soul}
\usepackage{pdfpages}

\usepackage{longtable}
\usepackage{makecell}
\usepackage{subcaption}

\def\HiLi{\leavevmode\rlap{\hbox to \hsize{\color{yellow!50}\leaders\hrule height .8\baselineskip depth .5ex\hfill}}}
         
\newcount\Comments  
\definecolor{darkgreen}{rgb}{0,0.6,0}         
         
\newcommand{\kibitz}[2]{\ifnum\Comments=1{\color{#1}{#2}}\fi}

\newcommand{\kg}[1]{\kibitz{red}{[Kobi says:#1]}}
\newcommand{\gb}[1]{\kibitz{brown}{[GB:#1]}}

\newcommand{\points}{\textsc{Points}}
\newcommand{\rank}{\textsc{Rank}}
\newcommand{\vfm}{\textsc{Vfm}}
\newcommand{\knap}{\textsc{Knap}}
\newcommand{\kapp}{\textsc{Kapp}}
\newcommand{\tapp}{\textsc{Tapp}}
\newcommand{\sw}{\textsc{sw}}
\newcommand{\voters}{{N}}
\newcommand{\mes}{ES}

\begin{document}

\title{Participatory Budgeting  Design for the Real World}
%
%

\author {
    Roy Fairstein\inst{1},
    Gerdus Benad\`e\inst{2},
    Kobi Gal\inst{1,3}
}

\institute{Ben-Gurion Univ. of the Negev, Israel \and
Boston University, USA \and
University of Edinburgh, U.K.}
\maketitle              
\begin{abstract}
Participatory budgeting   engages     the public in the process of allocating  public money to different types of projects. 
  PB designs  differ in how voters are asked to express their preferences over candidate projects and how these preferences are aggregated to determine which projects to fund. 
  This paper studies  two fundamental questions in PB design. Which voting format and aggregation method to use, and how to evaluate the outcomes of these design decisions?
  We conduct an extensive empirical study in which $1\,800$ participants vote in four participatory budgeting elections in a controlled setting to evaluate the practical effects of the choice of voting format and aggregation rule. 
We find that $k$-approval leads to the best user experience. 
With respect to the aggregation rule, greedy aggregation leads to outcomes that are  highly sensitive to   the input format used and the fraction of the population that participates. The method of equal shares, in contrast, leads to outcomes that are not sensitive to the type of voting format used, and these outcomes are remarkably stable even when the majority of the population does not participate in the election.  These results carry valuable insights for PB practitioners and social choice researchers.

\end{abstract}
%
%


\section{Introduction}\label{sec:intro}

Participatory budgeting (PB) is a type of direct democracy that allows a community to take part in deciding how to allocate a budget among different projects. It is growing in popularity and has been used around the world including in Madrid, Rome, Paris~\citep{sintomer2008participatory} and New York~\citep{su2017porto}.  
PB allows voters to affect local policies that matter to them, and  can increase community satisfaction when  better sets of projects are funded. \kg{does there exist work on satisfaction of public?}

Participatory budgeting elections consist of several phases. First, organizers elicit proposals for  projects. These proposals are developed further (expanded, merged, costed, etc.) and finally filtered to a set of candidate projects.  Citizens are then asked to cast a vote, and the votes are aggregated to determine which projects are funded. 
This paper assumes that a shortlist of candidate projects has already been fixed and studies the latter two components  of PB design: preference elicitation, or
how citizens vote  for their preferred projects,  and how votes are aggregated to determine the funded projects (aggregation). 
 
 PB  design is a complex task that   requires carefully balancing many factors. From the perspective of preference elicitation,   voters should be allowed to express complex preferences over projects.  For example, a voter may greatly prefer exactly one library being built rather than zero or two libraries, while being indifferent between the two proposed locations. At the same time, voting should be accessible to everyone in the community no matter their educational or socio-economic background, and there is a real risk that requiring too much information from voters may deter   participation. 
Very simple voting formats   come with their own problems and can lead to frustration and disillusionment. 
In practice,   participation rates in PB are often low  (in extreme cases as low as 0.1\% in Germany~\citep{zepic2017participatory} and 1-3\%   in Chicago in 2012 and 2014~\citep{stewart2014participatory,carroll2016democratizing}). 
Organising bodies  may also have secondary objectives, for example, designing an input format which exposes voters to   budgetary constraints similar to what the city council faces.

The vast majority of real-world PB elections use $k$-approval voting \cite{aziz2021participatory}, in which a voter approves their most preferred $k$ projects, though we are we aware of little formal justification for this choice.  Other formats that have been proposed in the literature or used in practice include knapsack voting (voters approve their most preferred budget-feasible set)  \cite{goel2019knapsack}, ranking projects by value or value-for-money  \cite{aziz2020expanding, benade2021preference} and reporting utilities \cite{peters2021proportional}.

When it comes to aggregation,  the funded projects should be justifiable given the observed votes and represent the preferences of  as large a part of the population as possible. 
Real-world elections almost exclusively use some form of greedy aggregation in which projects are selected in decreasing order of the number of approvals   received.  Alternatives including the Method of Equal Shares (\mes{})~\citep{peters2021proportional} , Cumulative Single Transferable Vote~\citep{skowron2020participatory} and variations on the greedy approach~\citep{talmon2019framework} have been proposed 
but are yet to be of practical significance.

\kg{does this problem statement come too late?}
This paper performs   a comprehensive  study of the effect of PB design on both user experience and the outcome of the election   towards answering the fundamental design question of
\emph{what input format and aggregation method should be used  in participatory budgeting elections?
}



We recruit more than $1\,800$ participants on Amazon Mechanical Turk,  present each with one of four different PB elections  and ask them to vote in one of six   input formats. 

We track several aspects of the voter experience, including the time it takes to learn and use each format and voters' self-reported belief that an input format accurately captures their preferences.  
We find that $k$-approval voting leads to the best voter experience overall. Voters using $k$-approval spent the least time  learning the format and casting their votes and found the format easiest to use. Somewhat surprisingly, voters also felt that  it was the format that allowed them to express their preferences best, despite the fact that $k$-approval captures strictly less information about voter preferences than, for example,  rankings over a set of projects. 

We compare greedy aggregation to \mes{}. Although the two methods lead to comparable social welfare, we identify two  advantages of \mes{} over greedy aggregation. First, it is much less sensitive to the input format used --- no matter the input format used, the set of projects that are funded remains almost identical. This makes \mes{} very versatile: as long as \mes{} is used for aggregation, the designer can choose almost any reasonable input format to satisfy whatever secondary objectives are present without fear that this will distort the outcome of the election. Second, we find that \mes{} is remarkably stable as participation decreases. In fact, sampling as few as 25-50\% of the voters and aggregating only the sampled votes rarely leads to a change in the set of projects being funded. In light of low levels of real-world participation, we believe this is a particularly attractive property (and one  not shared by greedy aggregation). 

Our study provides  a structured comparison of PB design choices with valuable insights for PB practitioners.  Our dataset and code will be made publicly available. 



\subsection{Related Work}\label{sec:related}
We briefly remark on the most closely related work; for a thorough review of the participatory budgeting literature see \citet{aziz2021participatory}.

Our work is closest to  \citet{benade2018efficiency}, who study the role of input formats in lab controlled PB settings. They asked participants  to vote over items that may aid their survival in a desert island. Each item has an associated weight and participants are informed that they are only able to carry items up to a specified weight limit. 
\citet{benade2018efficiency} report on the time it takes to vote, the consistency of preferences across format, and the distortion and welfare of different aggregation methods (under the assumption that the utility of an alternative is equal to the points assigned to it). 

In their work voters face only a single set of alternatives, and the desert island framing is far removed from PB as practiced in cities.  It is  hard  to argue that their findings are general  (for example, there may just be an obvious budget-feasible set of projects that best aids survival).   Our experiment much more closely mimics real-world participatory budgeting elections by using projects from real elections and  locating projects on a city map. We   attempt to check the robustness of our findings by repeating the experiment across four elections with different sizes and projects. 

\citet{goel2019knapsack} propose knapsack votes and empirically compare       value-for-money comparisons, knapsack and $k$-approval votes in several ways, including through a user survey. It is found that knapsack and $k$-approval votes take a similar amount of time, while value-for-money comparisons induce less cognitive load.  We find the opposite: voters consistently take significantly longer to  form value-for-money rankings. One possible explanation is that \citet{goel2019knapsack} performed a pair-wise comparison between projects, while we ask the voters for a full ranking over all projects. \citet{goel2019knapsack}   show some theoretical properties of knapsack under overlap utilities. 

\citet{benade2021preference}  study the theoretical information content of different input formats in a worst-case model.
They also conduct simulations, using votes from real PB elections,  to evaluate the social welfare that result from different input formats under a particular aggregation rule. They found that ``threshold approval" had both the strongest theoretical guarantees and highest welfare in simulations. 
Threshold approval does not stand out in our results.
%

A large literature studies the axiomatic properties of aggregation rules. Analysis most often focuses on maximizing the social welfare of the outcome \cite{benade2021preference,goel2019knapsack,jain2020participatory,hershkowitz2021district,talmon2019framework} \gb{other papers that maximize welfare?} 
or satisfying some version of proportionality, which (roughly) requires that a group of voters  with similar opinions should have impact on the outcome proportional to the size of the group \cite{fain2016core,  aziz2017justified, sanchez2017proportional, fain2018fair, aziz2018proportionally, skowron2020participatory, peters2021proportional} or both \cite{fairstein2022welfare,michorzewski2020price}.  Notably, \mes{} satisfies a proportionality condition called extended justified representation that guarantees a degree of representation to minority groups with common interests \citep{PS20}. 


We briefly look at welfare and representation but focus on evaluating  the stability of greedy aggregation and \mes{}, which  concerns the degree to which the outcome changes as one varies either the input format used in the election, or the degree of participation.  To the best of our knowledge this has not been studied before. 


\section{Preliminaries}\label{sec:prem}

For   $k\in \mathbb{N}$,  let $[k] := \{1,\ldots,k\}$. A PB instance is defined by a set of $n$ voters $N=[n]$ that express their preferences over a set of $m$  projects $P=[m]$ and a budget $B$. 
Each project  $p\in P$ has cost   $c(p)$. 
The purpose  of  a participatory budgeting process  is to select  a subset of projects $S\subseteq P$  to fund which satisfies the budget constraint, so $c(S) =  \sum_{p\in S} c(p) \leq B$.

We assume that voter $i$ gains utility $v_i(p)$ from project $p$ being funded. The value that  voter $i$ has for a set of projects $S\subseteq P$ is $v_i(S) = \sum_{p\in S} v_i(p)$, i.e. we assume additive utilities. 
The social welfare of project $p$ is $ v(p) = \sum_{i \in \voters} v_i(p)$ and the social welfare of outcome $S$ is $\sw(S) = \sum_{i\in \voters} v_i(S) = \sum_{p \in S} v(p).$ 

In reality, we do not have access to voters' true utilities. Instead, we receive a vote cast in some input format as proxy for these utilities. We study the following input formats.
\begin{enumerate}
    \item A \textit{utilities}  vote  (referred to as \points) asks a voter to divide 100 points between projects. Voter $i$ assigns $u_i(p) \in \mathbb{N}_0$ to each project $p\in P$ so that $\sum_{p\in P} u_i(p) =100. $ 
       
     \item  A \textit{$k$-approval}   vote (\kapp) asks a voter to approve (up to) their  $k$ most preferred projects. The $k$-approval vote of voter $i$ is represented as a binary vector $\alpha_i\in\{0,1\}^m$ with $\sum_{p\in P}\alpha_i(p)\leq k$.

    \item A \textit{threshold approval} vote  (\tapp) with threshold $t$ asks a voter to approve projects for which they have utility  at least $t$.  Voter $i$'s preferences are represented by a  binary vector $\alpha_i\in\{0,1\}^m$. 
    
    \item A \textit{knapsack} vote  (\knap)   asks each voter to select a subset of projects to maximize their utility subject to the budget. Voter $i$'s vote is a binary vector $\alpha_i\in\{0,1\}^m$ satisfying $\sum_{p\in P}  \alpha_i(p) \cdot c(p) \leq B$. 
    
    \item A \textit{ranking by value} (\rank)  is a strict total order over the projects. Voter $i$'s ranking is denoted $\sigma_i$, where  $\sigma_i(p)$ is the position of project $p$ in $i$'s ranking.  Observing   $\sigma_i(p_i) < \sigma_i( p_j)$    implies $v_i(p_i) \geq v_i(p_j)$. 

    \item A \textit{ranking by value-for-money} (\vfm) is a ranking of projects by the ratio of utility to cost, or `bang-for-buck'.  Voter $i$'s value-for-money ranking is denoted $\sigma_i$. Observing   $\sigma_i(p_i) < \sigma_i( p_j)$    implies $v_i(p_i)/c(p_i) \geq v_i(p_j)/c(p_j)$.
\end{enumerate}

The \knap{} and \vfm{} input formats require voters to know the budget and/or project costs.   \tapp{}  requires that voter utilities are normalized to the same scale.  


 
 Since we do not have access to true utilities  we deduce proxy valuations for projects  purely from the input profiles as follows:
For \points{}, we set $v_i(p) = u_i(p).$ For \knap{}, \kapp{} and \tapp{}, where the vote is a binary vector, we assume $\{0,1\}$
 utilities with $v_i(p) = \alpha_i(p)$. 
 For \rank{} and \vfm{}, we use Borda scores, so $v_i(p) = m - \sigma_i(p)$. 

After voters express their preferences, an aggregation method maps the input profile to a budget-feasible set of projects to fund. 
The first aggregation method that we consider is the \emph{greedy method}, which is most commonly used in real-world PB elections. 
It selects projects in  decreasing order 
of $v(p)$ until the budget is depleted, skipping projects as necessary.

The  second method we consider is called equal shares~\citep{PS20}, 
 or \mes{}. 
\mes{} virtually  divides the PB budget equally among all the voters. 
At each iteration, it identifies the remaining projects which can be fully funded by their supporters using their remaining virtual funds.
It then funds from among these the  project $p$ with the smallest ratio between $c(p)$ and $v(p)$. 
The cost of this project is divided among its supporters proportional to their utility. 
This process terminates when no project can be funded by only its supporters. 
%
One quirk of \mes{} is that it may not return a set of projects which exhaust the budget. To handle this case, we fill out the selected subset by perturbing voter values so that each voter has non-zero value for every project, as proposed in  \citet{peters2021proportional}.


\section{User Study Description}\label{sec:description}


The user study consists of asking voters to vote using one of the six input formats above in one of four different participatory budgeting elections in a hypothetical city. 
We recruit roughly 75  different participants for each of the 24 configurations (four elections times six input formats)   using Amazon Mechanical Turk, in total  just over $1\,800$ participants.\footnote{The data can be found at\\ https://github.com/rfire01/Participatory-Budgeting-Experiment} Participants' demographic characteristics are summarized in the appendix.   

The small elections (\textsc{Small-A} and  \textsc{Small-B}) consisted of 10 candidate projects while the larger elections  (\textsc{Large-A},  \textsc{Large-B}) had 20 projects, all with a budget of $500,000$.
Project descriptions and costs were taken from real participatory budgeting elections from across the world. \gb{I removed the footnote - doesn't make sense to just giev 1. }  
Each project was assigned a location on a map of the  city  and categorized in one of five  categories   (education; streets and transportation; culture and community; facilities and recreation; environment, public health and safety). 
Full  details  on every election including the project titles, descriptions,  costs and locations may be found in the appendix.

 \begin{figure*}[h]
\begin{center}
\includegraphics[clip, trim=0 3mm 0 0, width=13cm]{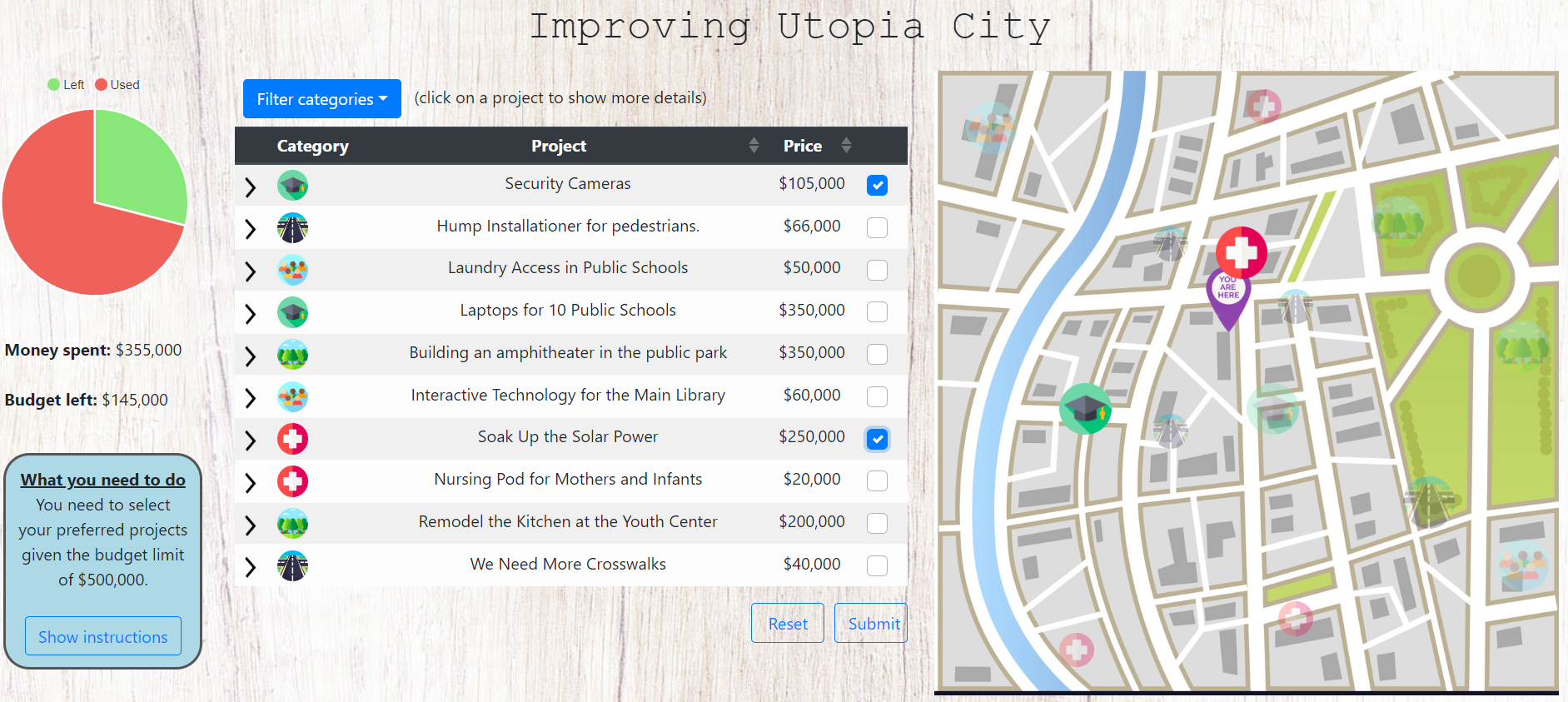}
\caption{The GUI shown to participants using the \knap{}  input format (center), showing  the city map (right) and budget (left)
}\label{fig:interface}
\end{center}
\end{figure*}


Participants were first presented with a written and video description of the PB voting task. They had to pass a simple quiz about the task in order to proceed. 
Next, participants carried out the voting task 
in their allocated PB configuration.
Each participant was assigned a  (random) location on the city map and   shown the description and location of the projects. 
We expected that a voter's location relative to  the project may affect their   preferences (e.g., people may prefer to vote for a library  close to their current location).


 Figure~\ref{fig:interface} shows the interface  presented to participants who were asked to cast knapsack votes. 
 The left-most column shows the instructions and an indication of what fraction of the budget remains to be allocated. The center column shows the category and title of each proposed project, and a participant can click on a project to see a more detailed description. The right-most panel shows a map of the city on which the locations of the voter and the projects are marked (hovering over a project highlights its location). The other interfaces used in the study may be viewed in the appendix. Project costs was only displayed for \points{} and \vfm{}, where it is needed to form a vote. 



 
 After   submitting their PB vote, participants were required to answer several consistency questions about their vote that were  designed to identify voters who vote randomly or carelessly (exact questions can be found in appendix).  Finally, participants were asked to complete a short survey about their subjective experience of voting in the assigned format. They were asked to rate (on a scale of 1 to 5) how easy they found the task, how much they liked the user interface, 
  how   expressive  they found the input format, 
 and how much the project categories and their location on the map    affected their decisions (exact questions can be found in appendix). 
 
 Participants were rewarded a  fixed sum    for participation and  received a 75\% bonus for passing the  consistency questions.
 IRB approval was obtained from the corresponding institution.


\section{Effect of Input Format on Voter Experience}

We investigate the practical effects of using different input formats by comparing  the experiences of voters using the respective input formats.

This comparison has several components. Objectively, we record the time that it takes a voter to complete each of the first three stages of the task (completing the tutorial, answering the post-tutorial quiz, and casting a vote). 
We also report the number  of participants who fail the post-task consistency test and argue that a participant's inability  to recall simple information about the vote they just cast is correlated with the participant finding the task taxing, confusing or tiresome. 
More subjectively, we examine participants' self-reported scores from the post-completion survey. 




\subsection{Response time}
\begin{figure}[!h]
\begin{center}
\includegraphics[width=8.5cm]{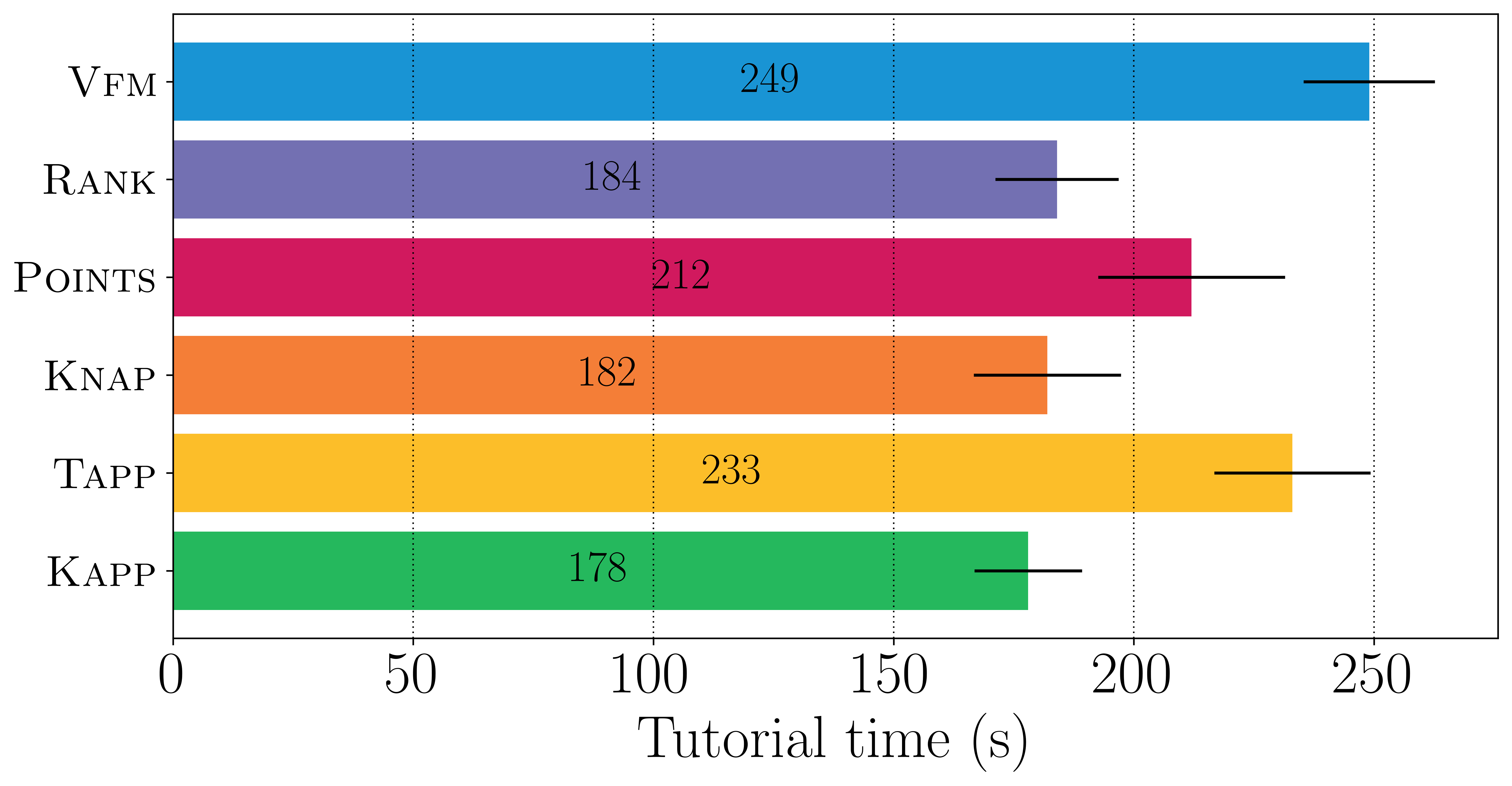}
\includegraphics[width=8.5cm]{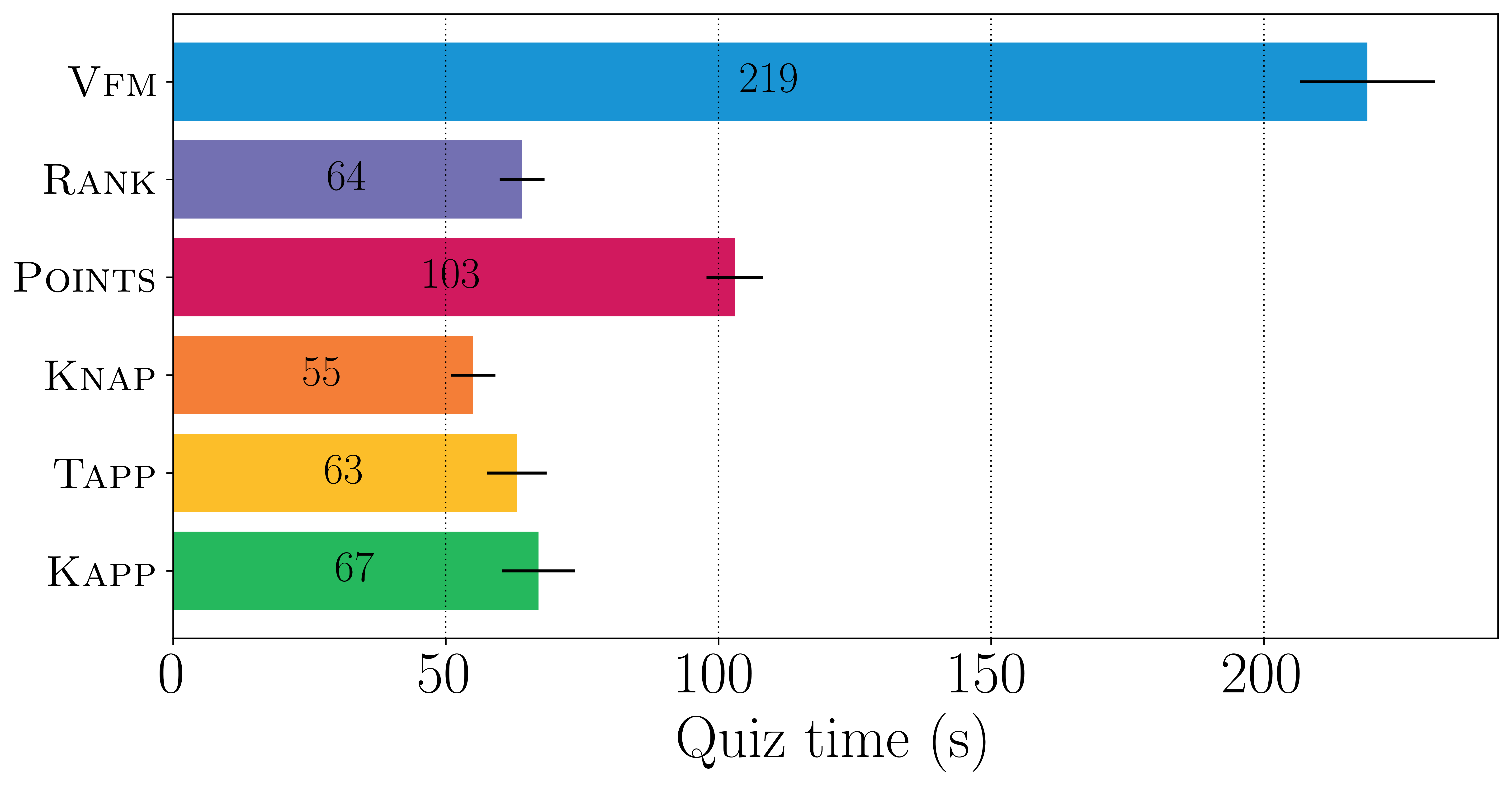}
\includegraphics[width=8.5cm]{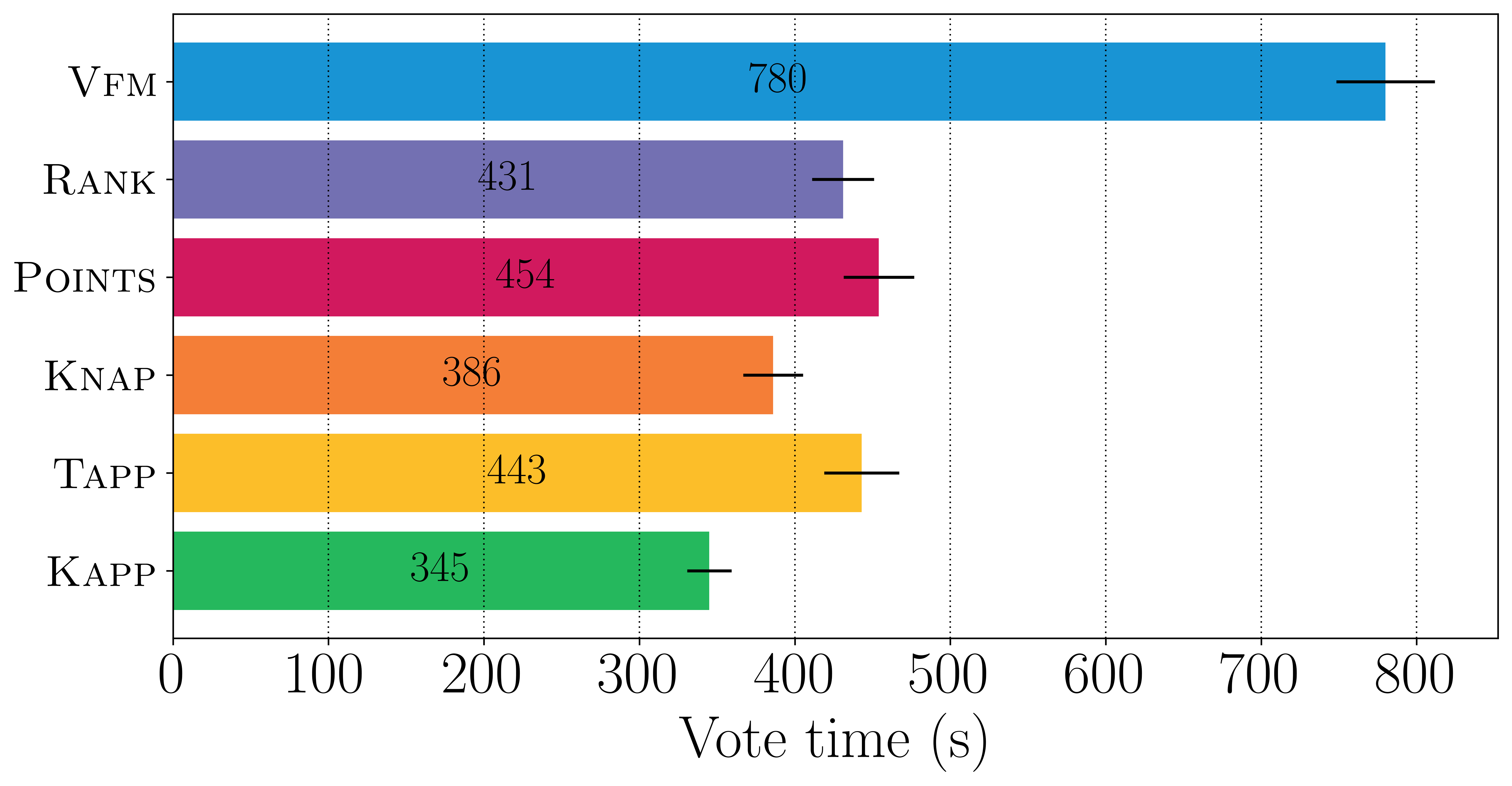}
\caption{Average time ($s$) to complete each stage. 
}\label{fig:time}
\end{center}
\end{figure}

The time it takes to complete a task is a recognized  proxy for  the cognitive burden  or difficulty of the task  \cite{rauterberg1992method}.  
The average time to complete each stage of the experiment is visually represented in  Figure~\ref{fig:time}.  Results in this section are averaged across elections and the conclusions do not change when looking at any individual election. 

Participants voting in  the \vfm{}   format are consistently slowest to complete each stage of the experiment, suggesting that voters find \vfm{}  very hard to use. Based on the time it takes to cast a vote,  \points{} and \tapp{} are the next hardest formats to use. In the former case, we believe this supports the idea that asking voters for cardinal utilities is generally not feasible; in the latter, it  highlights the complications that sprout from  \tapp{} requiring cross-voter normalization  (as well as the fact that \tapp{} is a relatively niche format which voters are unlikely to have encountered before). 


Voters found \kapp{} the easiest to use, followed by \knap{} and \rank{}.

\subsection{Consistency checks}

Of the roughly 75 participants  recruited for each of the 24 configurations of the experiment, on average 58 passed the consistency test. 

The consistency test  is designed to check whether the participant can recall simple information about the recently completed task. 
There are several reasons why a voter may fail the test. For example, they may answer randomly in order to complete task quickly,  become  distracted  if the task is too tiresome, or confused if the input format is hard to understand. As such, we  treat this as another signal about the difficulty of expressing  preferences in a given format.




We  compare the rate of passing the consistency test across input formats. 
Participants using the ranking-based formats were most consistent, followed by the approval-based formats.  We speculate that the high consistency of \vfm{} is, at least in part, thanks to the inordinately long amount of time voters spend considering their votes. Users of \points{} were comfortably the least consistent, which  supports earlier evidence that providing cardinal utilities is a challenging task.

There was virtually no variation in the rate of passing the consistency test across the four elections. A complete breakdown of the results may be found in the appendix. 


\subsection{Self-reported voting experience}
\begin{figure}[!h]
\begin{center}
\includegraphics[width=10cm]{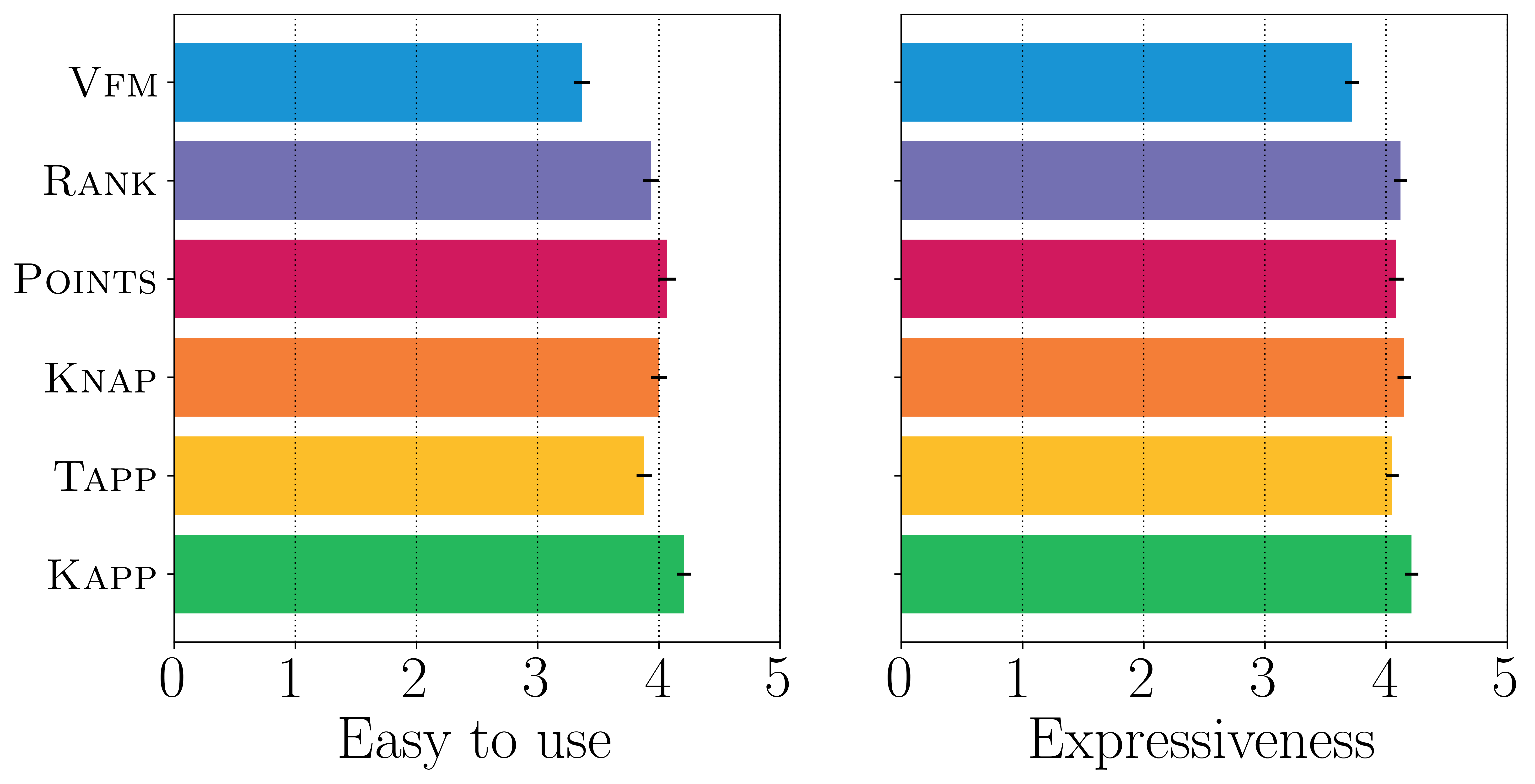}
\caption{The average feedback for each input format.
}\label{fig:feedback}
\end{center}
\end{figure}

In addition to the time it took participants to complete the task and how much they could recall about their submitted preferences, we also expressly asked them to rate various aspects of the experience on a scale of 1 to 5. 
Survey results are summarised in  Figures~\ref{fig:feedback} and \ref{fig:cat_map}.
Results for the questions omitted from this discussion may be found in the appendix.
In order to find statistical significance, we first used Kruskal–Wallis test to show the input formats give different results, followed by the Dunn test for post hoc comparison.
All of the survey questions   gave $p < 0.004$  for the Kruskal test. The Dunn test evaluated statistically significance   at the $p < 0.05$ level.

\textbf{Ease of use}  We asked the voters how easy they found the voting task. Unsurprisingly, participants found \vfm{} significantly harder to use than any other format. \points{} was rated as quite easy to use despite being one of the more time-intensive formats. \kapp{} was rated as significantly easier to use than all other  formats except \points. 


\textbf{Perceived expressiveness} We asked the participants how well they felt their vote captured their preferences. 
Participants found \kapp{} to be most expressive and, in particular,  significantly more expressive than  \tapp, \vfm, and \points. It is somewhat surprising that \kapp{} was rated as more expressive than \points, since you can infer a \kapp-vote from  your \points-vote; we speculate that the relative complexity of  \points{} explains part of this phenomenon. It is also possible that participants conflated expressiveness and ease of use, or failed to consider alternatives to the format presented to them. 
\vfm{} was perceived to be least expressive by some margin. Again, the  difficulty of using \vfm{} and the fact that voters are forced to explicitly consider project costs may have contributed to the perception that it is less expressive than \rank, which also asks for a ranking of alternatives.

\begin{figure}[!h]
\begin{center}
\includegraphics[width=10cm]{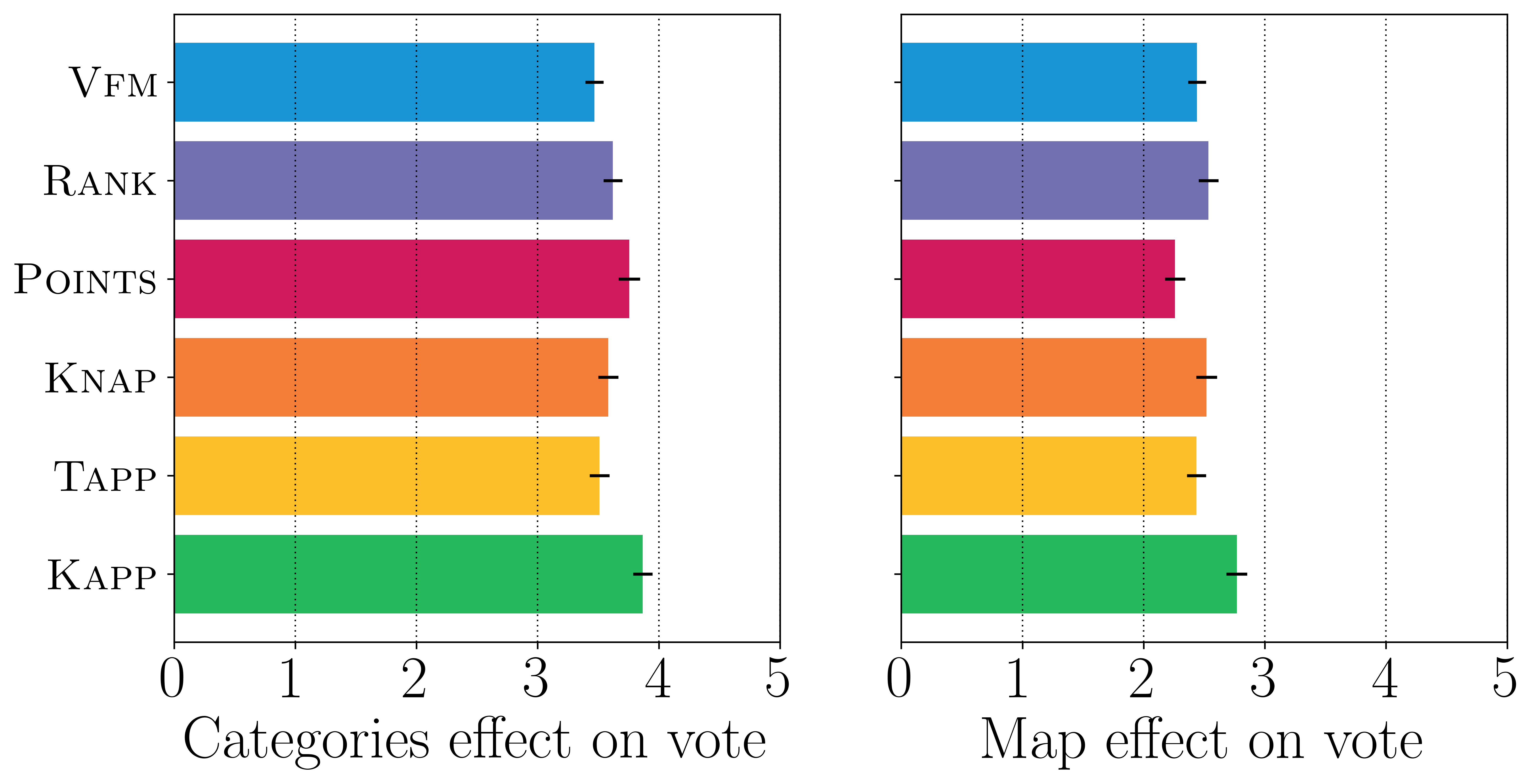}
\caption{The average effect of the categories and the map.
}\label{fig:cat_map}
\end{center}
\end{figure}


\textbf{Effect of additional information}
Projects were assigned to one of five categories (e.g.\ transport or education) and given locations on a city map. Two survey questions asked about the extent that project categories and the map influenced participants' preferences. These results are summarised in Figure~\ref{fig:cat_map}

Participants using \kapp{} were influenced  more by project locations than any other group. Similarly, \kapp-voters paid significantly more attention to project categories than all other groups except \points-voters. 
It may be that the \kapp{} format was easy enough to understand and use that it freed participants up to consider additional, non-core information about the projects and   election in their decisions. 



\section{  Effect of Aggregation Method on  Outcome}

\label{sec:aggregation}
In  this section we  analyze the effect of aggregation methods  on the outcome of the elections.

\subsection{Stability} 




We first study the   stability of the different aggregation methods. 
We say that an aggregation method is stable when the outcome of the election is:
\begin{enumerate}
\item Robust to the choice of input format. This allows the organizers to freely select an input format without fear of affecting the outcome of the election (or default to a format voters find easy to use). 
\item  Robust to partial participation.   Voter turnout may be low in real PB elections, ideally the outcome of the election is not drastically affected by the (non-)participation of a small group of voters. 
\end{enumerate}

In this experiment, we first repeatedly sample $n'=40$ participants per configuration, and aggregate the resulting vote profiles using greedy aggregation and \mes{}. This process is repeated 200 times.

\begin{figure*}[h]
\begin{center}
\includegraphics[width=14cm]{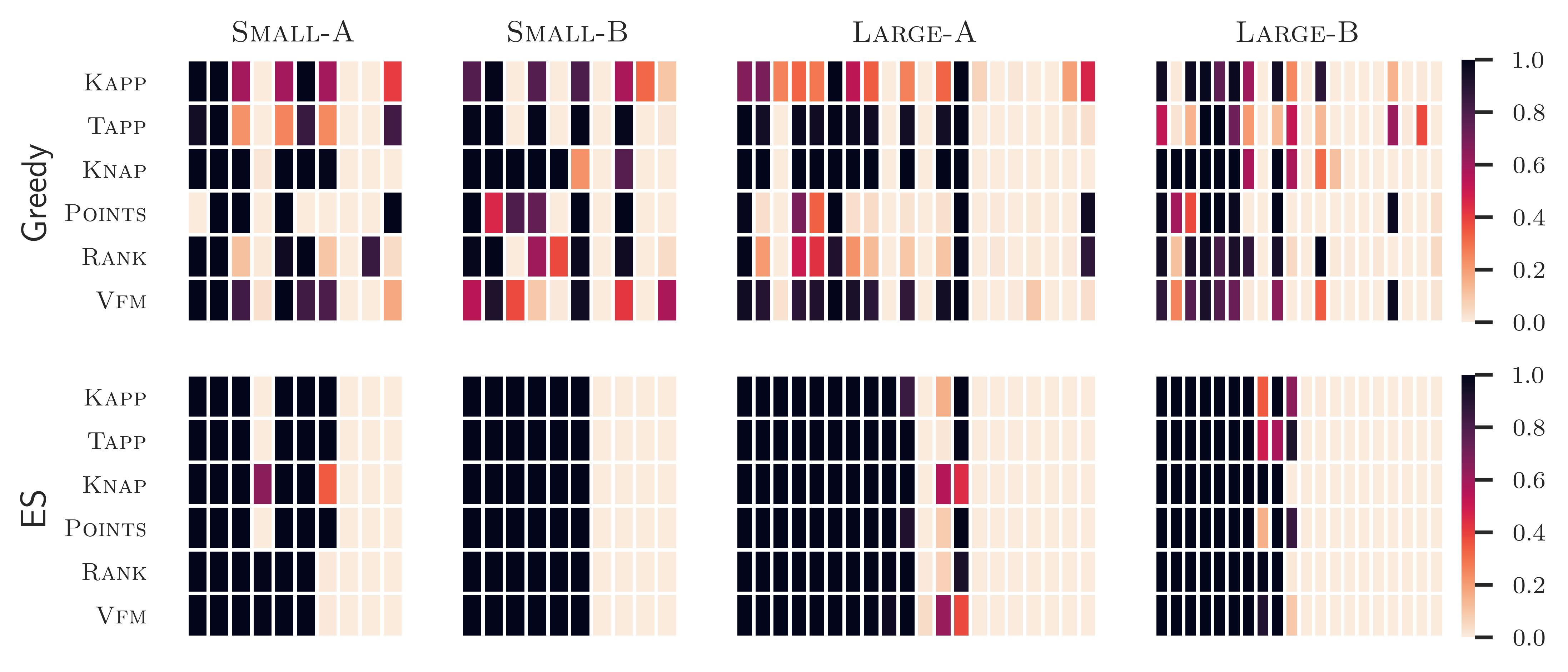}
\caption{Stability heatmaps for  greedy aggregation (top) and \mes{} (bottom) in each election. 
Within each panel each row represents an   input format and each column a project (ordered in increasing cost).
 The intensity of a cell indicates the fraction of instances in which the project was funded.
}\label{fig:heatmap}
\end{center}
\end{figure*}

The fraction of repetitions in which each project is funded in each configuration is summarized in a heatmap in Figure~\ref{fig:heatmap}. The top row of results corresponds to greedy aggregation and the bottom row to \mes. Within each panel projects are ordered in order of increasing cost, and the cells range from white (meaning the project is never funded) to black (always funded).



Strikingly, the outcome of the election is almost entirely unaffected by the choice of input format when using \mes{} --- the majority of the projects are either always funded under every input format or never under any. Greedy aggregation exhibits this property only in rare instances, for example, the outcome of \textsc{Large-A} does not change if the input format is changed from \knap{} to \tapp.  This robustness to the choice of input format remains present when the number of sampled voters is varied $n'\in\{10,20,30\}.$

We also observe in   Figure~\ref{fig:heatmap} that \mes{} is robust to partial participation: it is very rare that the decision to fund a project changes across repetitions when sampling voters at random, i.e.\ under a uniform model of voter participation/abstention. Greedy aggregation does not exhibit this property. 

Of course, since we are sampling 40 votes from (on average) 58 consistent voters the sampled profiles are correlated. We repeat the experiment sampling $n'\in\{10,20,30\}$ voters each time to investigate what happens when the correlation in sampled profiles decreases.
To quantify stability under partial participation, we compute the entropy of the outcome for each configuration. Suppose that project $p\in P$ is funded with probability $f_{V,A}(p)$ when aggregating votes cast in voting format $V$ using aggregation method $A$, then $\text{entropy}(V,A) = -\frac{1}{|P|}\sum_{p\in P} f_{V,A}(p) \cdot \log_2(f_{V,A}(p)) + (1 - f_{V,A}(p)) \cdot \log_2(1 - f_{V,A}(p))$.

\begin{figure}[!h]
\begin{center}
\includegraphics[width=7.5cm]{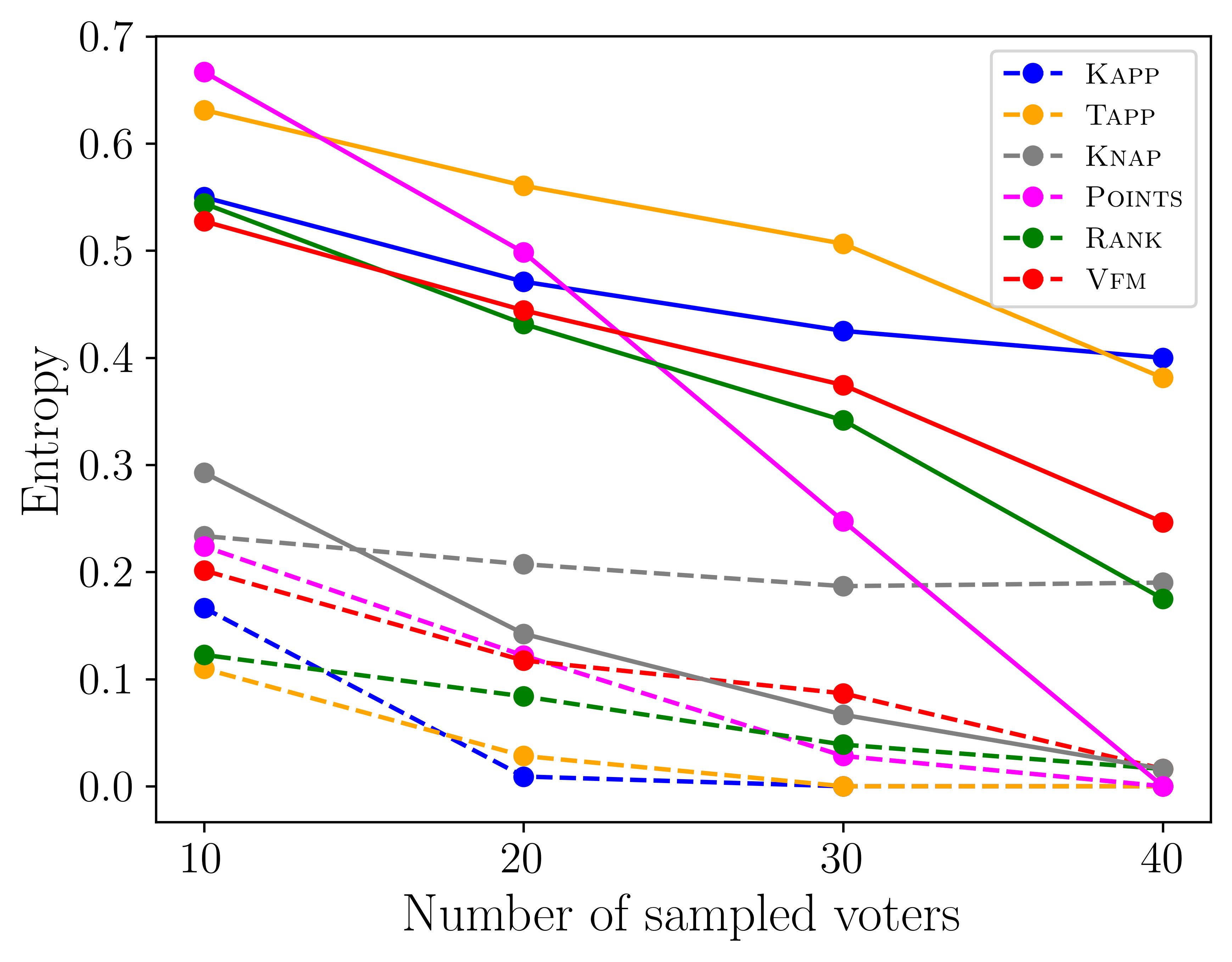}
\caption{Entropy for each input format under  greedy aggregation  (solid lines) and \mes{} (dotted lines)  as the number of sampled voters changes. }\label{fig:entropy}
\end{center}
\vspace{-3mm}
\end{figure}

Figure~\ref{fig:entropy} shows the entropy for election   \textsc{Small-A} across   input formats for greedy aggregation and  \mes{} aggregation when varying the degree of participation (i.e. the number of sampled voters). 
Unsurprisingly, the correlation between vote profiles decreases and entropy increases  as fewer voters are sampled. 
The entropy of \mes{} is consistently significantly lower than that of greedy aggregation across input formats. We conclude that \mes{} is significantly more robust to partial participation than greedy aggregation. One exception is worth highlighting, the entropy of greedy aggregation with \knap{} votes is fairly competitive with that of \mes. This suggests that if the organizers of a PB election are dead-set on using greedy aggregation and also worried about the effect of partial participation,  \knap{} may be an attractive option.  
%
For the other three elections,   \mes{} aggregation comfortably outperforms greedy aggregation across all input formats and sample sizes (except similar results for \knap{} in election \textsc{Large-A}). Full results  may be found in the appendix.


We conclude this section with an observation from  Figure~\ref{fig:heatmap} unrelated to stability. It appears to be the case that more expensive projects are rarely funded  when using \mes{}.  We believe this may (at least for the binary input formats)  be an artefact of assuming $v_i(p)\in \{0,1\}$, which makes it very hard for an expensive project to have a ratio of utility to cost that justifies funding it. It remains to be seen whether this trend persists, for example, when assuming $v_i(p)\in \{0,c(p)\}$.


\subsection{Social Welfare}
An obstacle to comparing outcomes based on social welfare is participants are assigned only one format, so there is no common basis of comparison when aggregating vote profiles from different formats. To address this, we make the assumption that, since voters are randomly assigned to input formats, the distribution of voter preferences are the same across formats. As such, we can deduce voter valuation functions from the full vote profile in \emph{any} format and treat these aggregated values as a proxy for the welfare of all voters, regardless which format they used.\footnote{ We caution that this is at best a very coarse evaluation, since assuming a value based on  a vote in any of the formats requires very strong assumptions. 
} 

\begin{figure}[!t]
\begin{center}
\includegraphics[width=6cm]{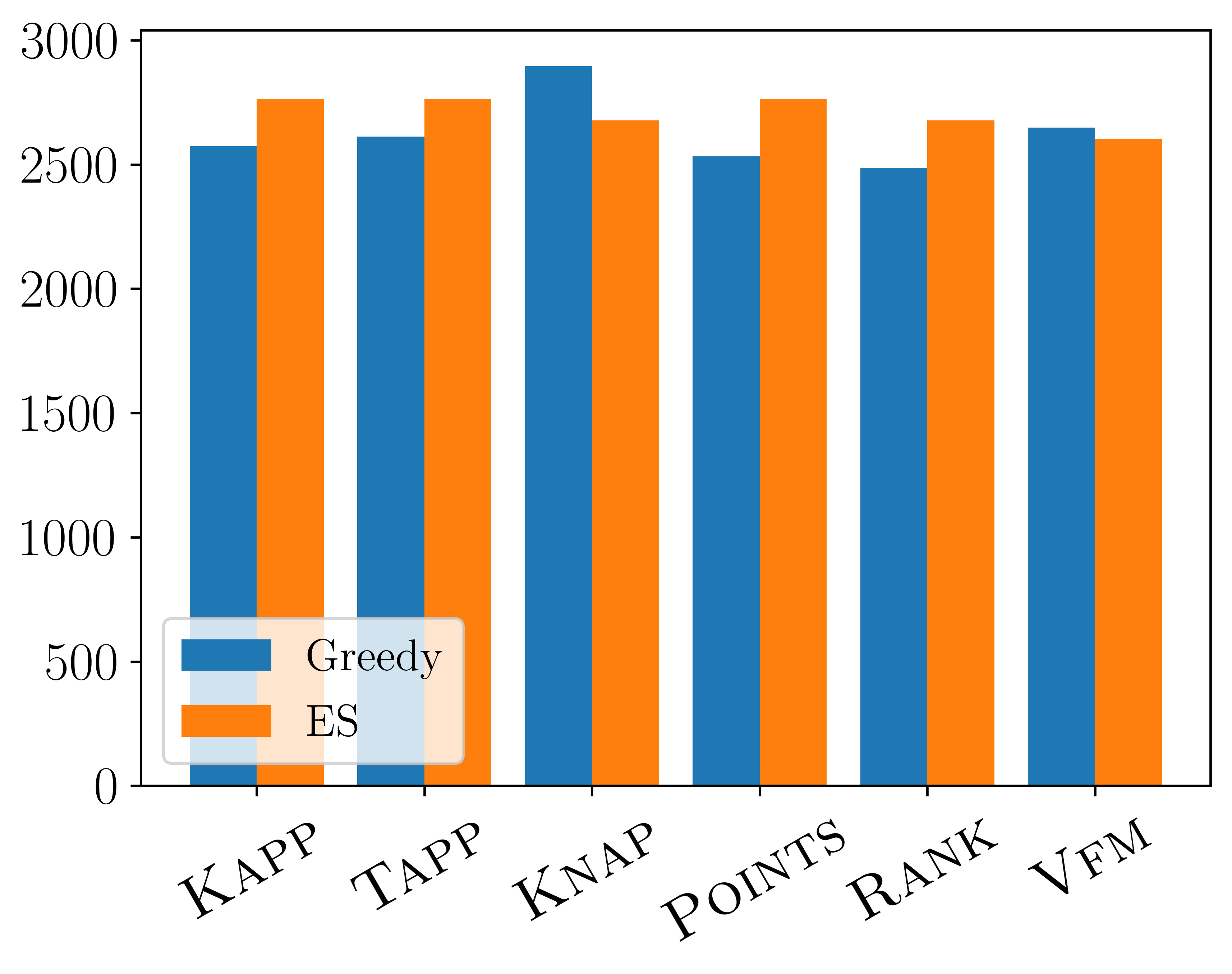}
\includegraphics[width=6cm]{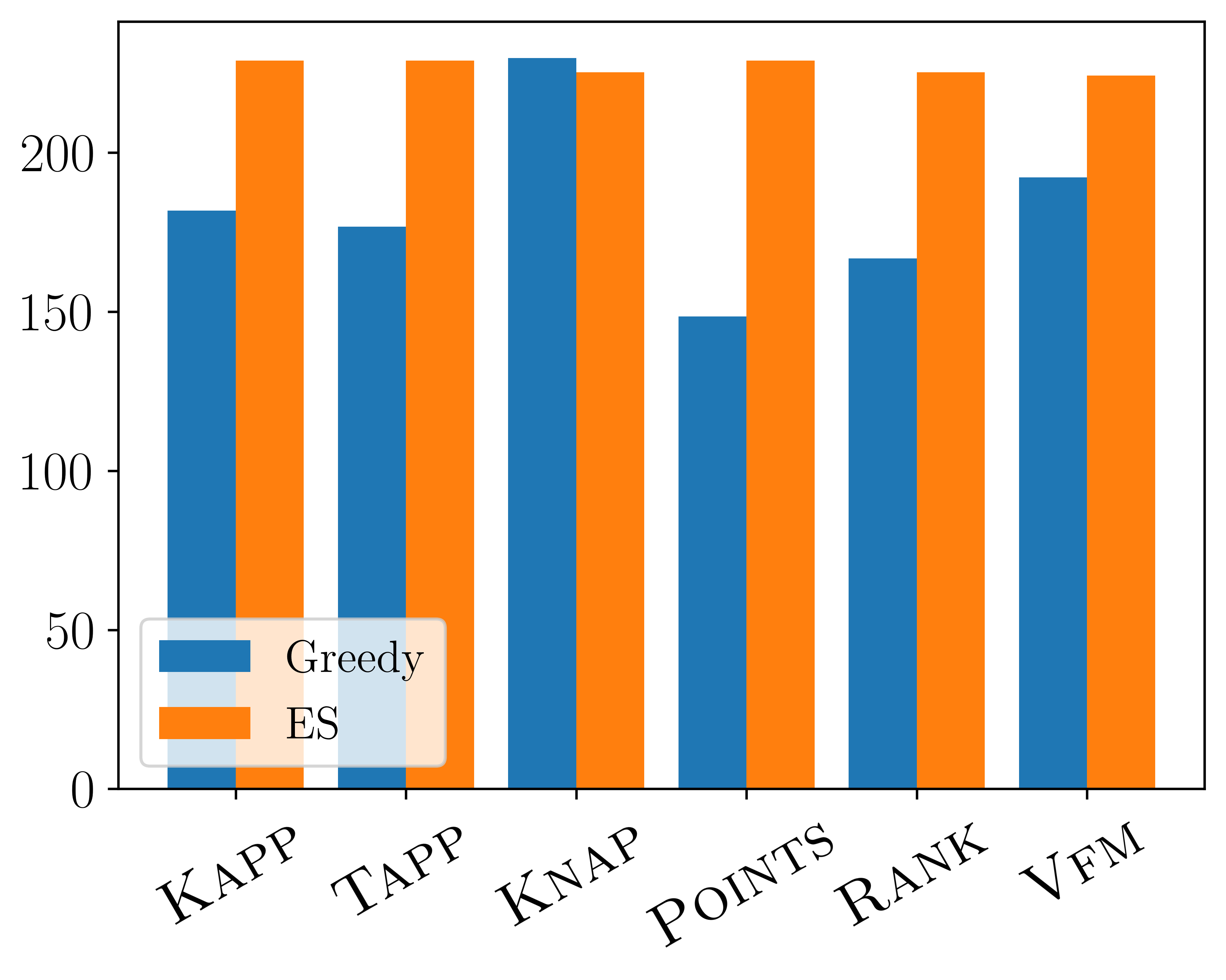}

\caption{Average social welfare (across elections) using the \points- (left) and \knap-voters (right) to compute welfare.
}\label{fig:welfare}
\end{center}
\end{figure}

We report in Figure~\ref{fig:welfare} the   social welfare (averaged over the four elections) as measured when treating the full profile of \points{} or \knap{} voters as representative.  


Neither aggregation method strictly outperforms the other. However, we observe that, when using greedy aggregation, the choice of input format can have a large effect on welfare. This is seen most clearly when using \knap-voters to compute welfare. Unsurprisingly, \mes{} is less sensitive to  the choice of input format and   achieves very similar welfare across input formats.
Results when using the other formats to measure welfare are less extreme than the results shown here and may be found in the appendix.




\section{Discussion}

These results highlight practical insights  for the design of real participatory budgeting elections. 
When choosing an input format, we find no reason to deviate from the current standard practice of  using \kapp{} voting: Voters find the format easy to understand, easy to  use, and believe it allows them to express their preferences accurately. 

Greedy aggregation is used almost universally,  however, our experiments find that \mes{} has  two  big advantages over greedy aggregation.  First,  the outcomes are largely stable and unchanged when only a subset of the population participates. In light of low levels of real-world participation, we believe this is a particularly attractive property. Second, \mes{} appears to be robust to the choice of input format, in other words, it affords the city officials responsible for administering the participatory budgeting election a great deal of freedom to choose an input format which meets whatever secondary objectives are deemed to be important. 

One argument that remains in favour of greedy aggregation is its transparency: it is arguably easier to explain to voters than \mes{}. If greedy aggregation is used    and stability to partial participation is a primary concern, then \knap{} emerges as a fairly user-friendly format overall and a clear front-runner in terms of stability.

These recommendations should be taken in the context of the limitations of this study. First, the  behavior and preferences of participants recruited on Amazon Mechanical Turk may not be sufficiently similar to those of real voters.  Though we attempt to  ensure the findings are robust across the parameters of different elections by varying the number and type of projects, there is no mimicking the richness, variety and local idiosyncrasies of real-world instances. Similarly, real voter utilities likely exhibit complementarities and externalities  --- a far cry from our utility proxies. Second, it is possible that, despite our best efforts, the   voter experience   was affected by the interface design choices we made. 

We conclude with  two directions for future study. 
When it comes to designing participatory budgeting elections, the proof, to a large extent, should be in the pudding.  
One can imagine a  variant of the experiment presented in this paper in which votes from different PB designs are aggregated and voters are directly asked to compare different elections and outcomes on the transparency of the process, personal satisfaction, perceived fairness of the outcome, etc. 
Directly asking people their opinion about PB outcomes may yield new  insights   that can help inform design decisions while avoiding assumptions like  additive utilities. 

Democratic innovations live and die by the extent to which voters' voices are heard. But this requires representative participation. 
As a fledgling part of the democratic process, participatory budgeting appears to be particularly susceptible to poor participation. It is important to understand how   design choices around voting formats and the transparency of aggregation rules affects the participation (now and in the future) of voters from a wide range of socio-economic backgrounds. Our stability experiments are a step in this direction but assume participation is uniform across the population, which is unlikely to be the case in reality.

\section{Acknowledgments}
Nisarg Shah provided instrumental help and advice with all aspects of this research. Ariel Procaccia contributed to an earlier version of this paper. Thanks to Boaz Avami for programming assistance with the study.

%
%
%
\bibliographystyle{plain}

\bibliography{ref.bib}
%

\clearpage
\appendix


\section{Interface For Voting}\label{app:interfaces}
\begin{figure*}[ht!]
     \centering
          \begin{subfigure}[b]{0.45\textwidth}
         \centering
       \includegraphics[width=6cm]{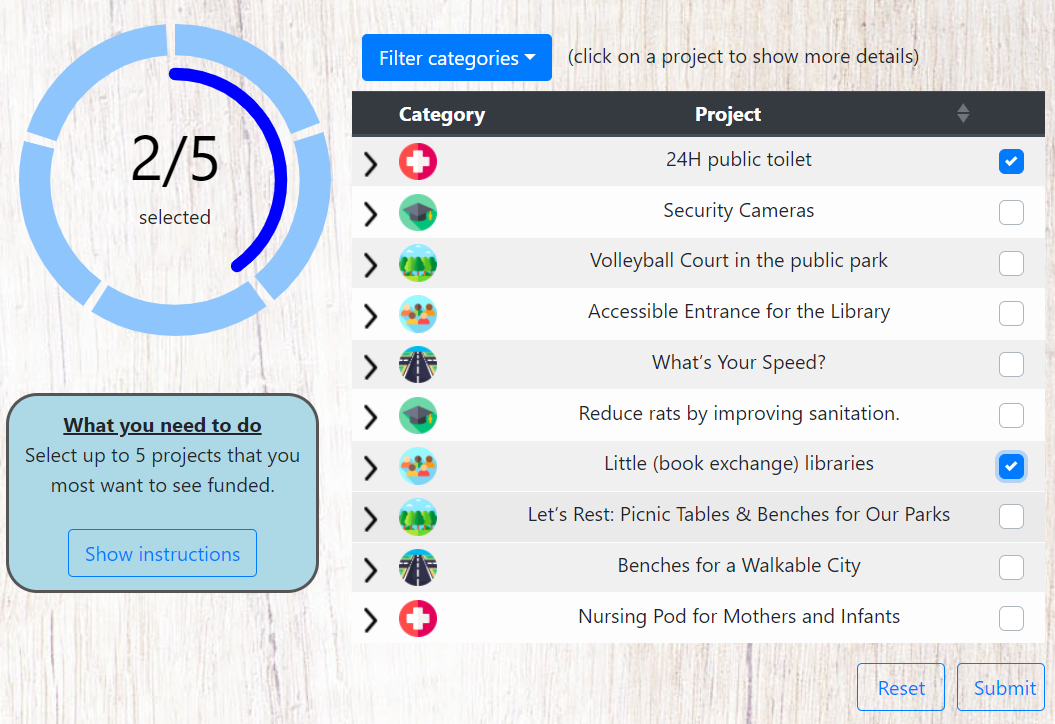}
\caption{\kapp{} user interface.
}\label{fig:kapp_inter}
     \end{subfigure}\hfill
     \begin{subfigure}[b]{0.45\textwidth}
         \centering
         \includegraphics[width=6cm]{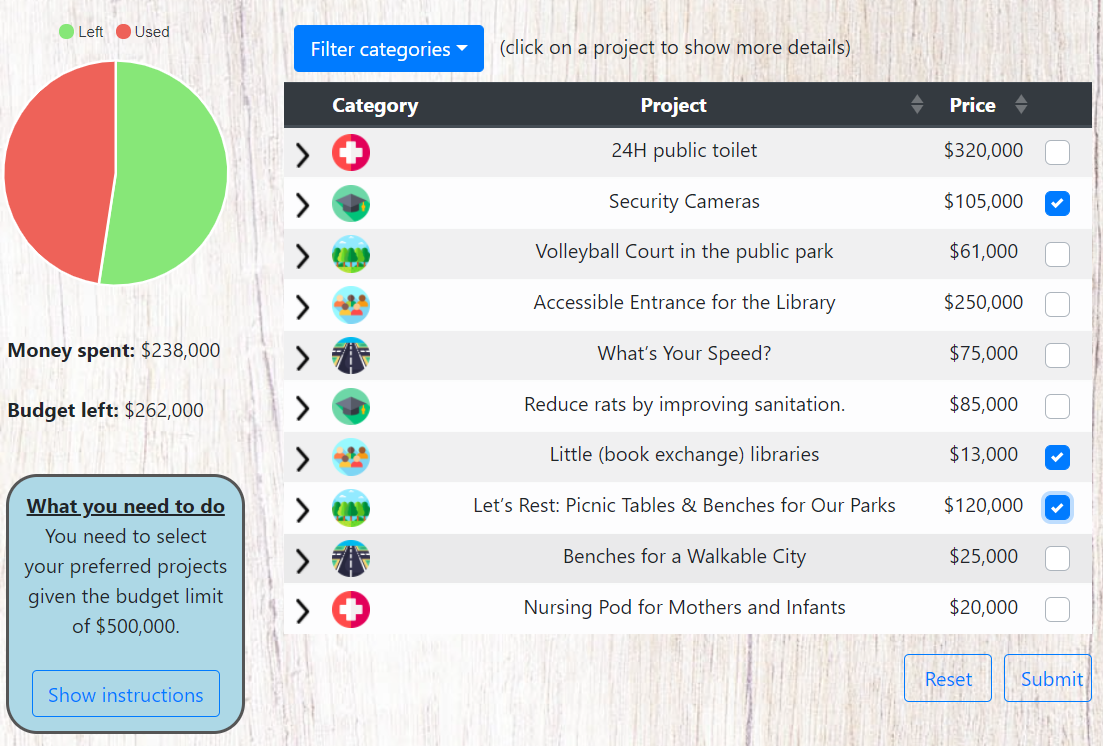}
\caption{\knap{}  user interface.
}\label{fig:knap_welfare_app}
     \end{subfigure}
     \hfill
     \begin{subfigure}[b]{0.45\textwidth}
         \centering
         \includegraphics[width=6cm]{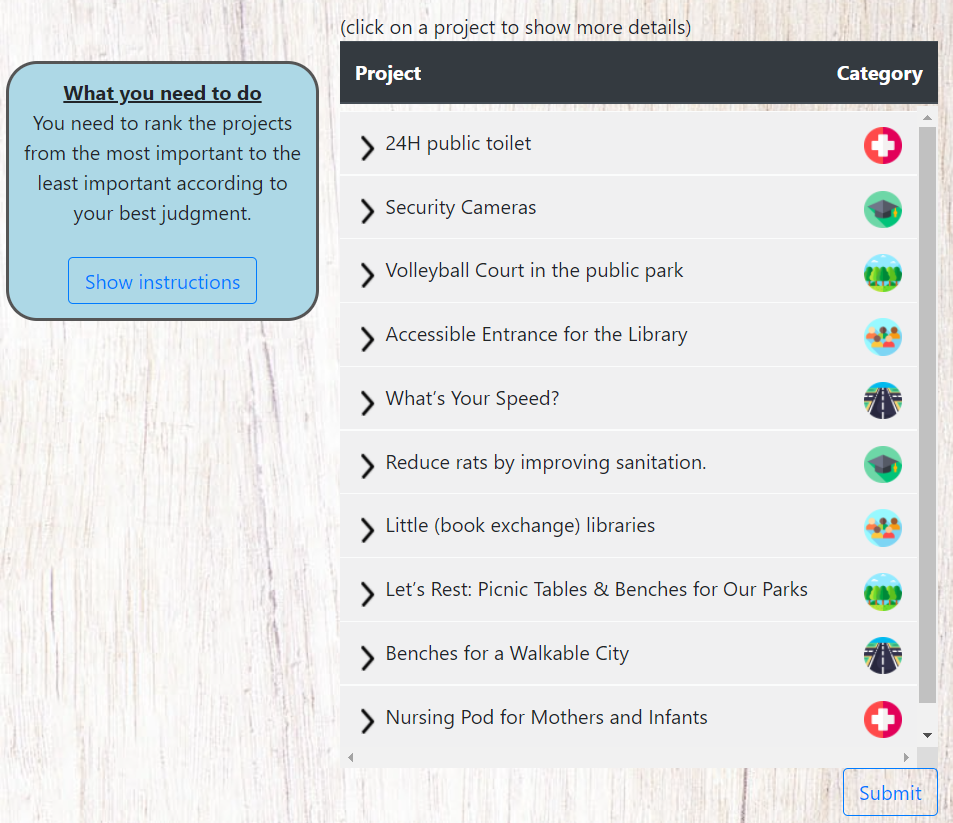}
\caption{\rank{} user interface.
}\label{fig:rank_inter}
     \end{subfigure}
     \hfill
     \begin{subfigure}[b]{0.45\textwidth}
         \centering
\includegraphics[width=6cm]{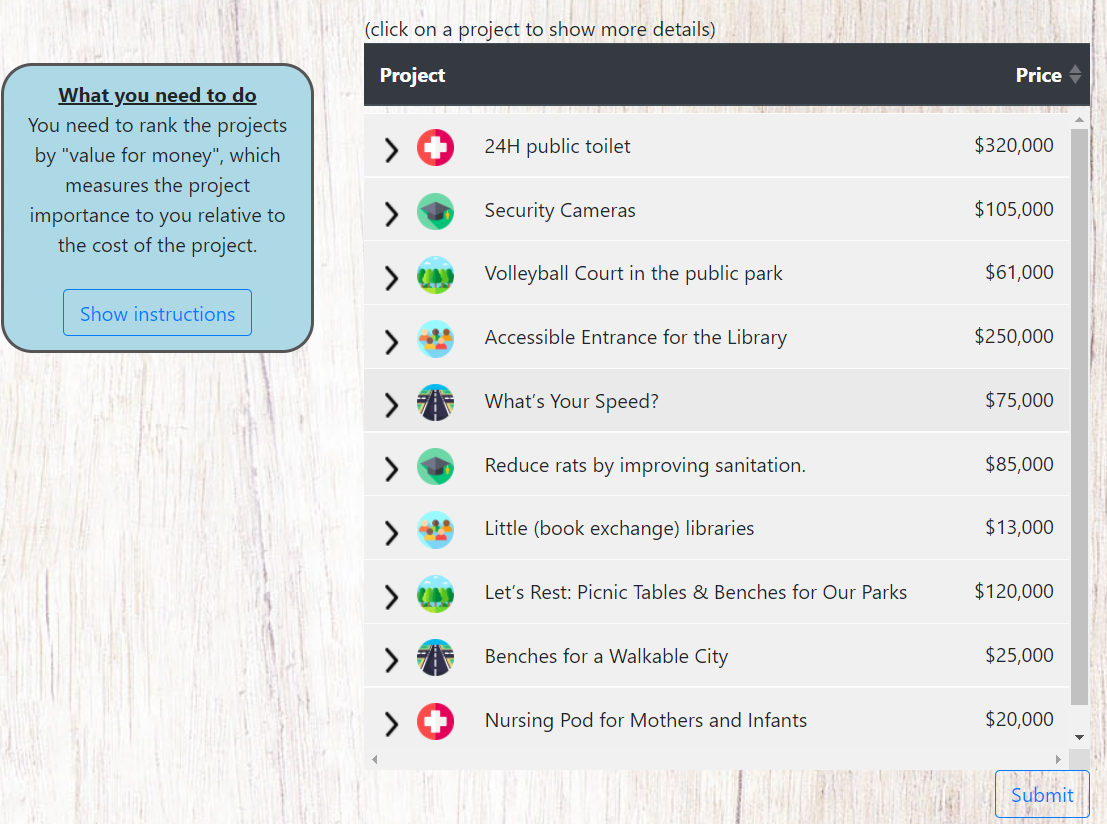}
\caption{\vfm{} user interface.
}\label{fig:vfm_inter}
     \end{subfigure}
     \begin{subfigure}[b]{0.45\textwidth}
         \centering
\includegraphics[width=6cm]{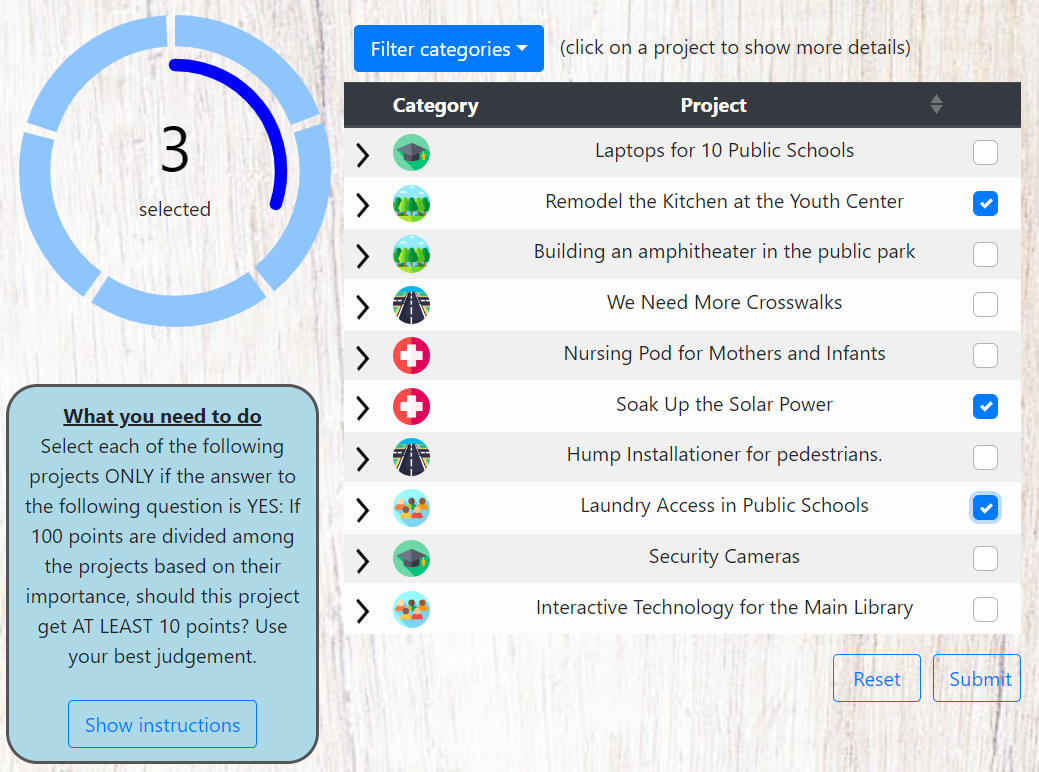}
\caption{\tapp{} user interface.
}\label{fig:tapp_inter}
     \end{subfigure}
     \hfill
     \begin{subfigure}[b]{0.45\textwidth}
         \centering
\includegraphics[width=6cm]{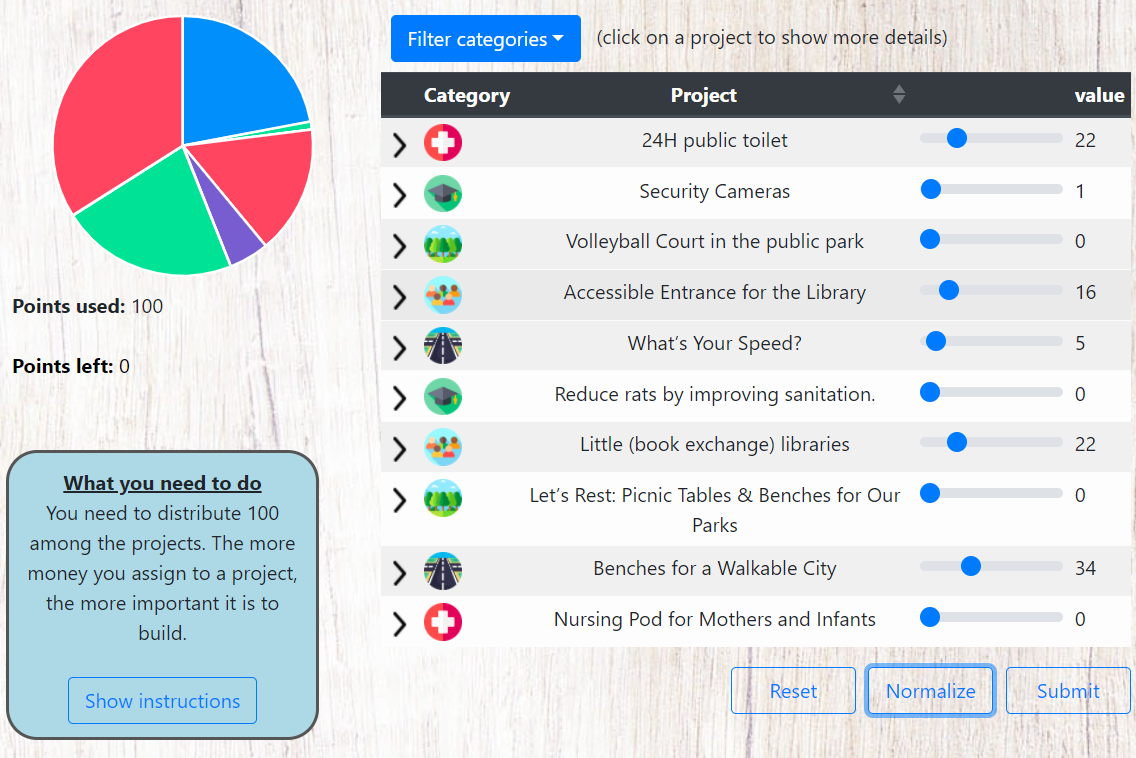}
\caption{\points{} user interface.
}\label{fig:util_inter}
     \end{subfigure}

        \caption{Input formats interface.}
        \label{fig:all_interfaces}
\end{figure*}

Figure~\ref{fig:kapp_inter}-\ref{fig:util_inter} shows   the interface for each of the input formats.

\section{Experiment Configurations}\label{app:elections}

In the experiments each participant encountered one of six input formats and one of 4 possible elections, each with different set of projects, either 10 projects or 20, such that each project belong to one of 5 categories:

\begin{itemize}
    \item Culture and community.
    \item Streets, Sidewalks and Transit.
    \item Environment, public health and safety.
    \item Facilities, parks and recreation.
    \item Education.
\end{itemize}

Tables~\ref{tab:first_elc}-\ref{tab:fourth_elc} show the four different elections the participants might encounter in the experiment, a short description, and to which category it belongs. The given coordinates are for 100x100 map, certain projects include several coordinates.

\begin{longtable}[ht!]{|p{2cm}|p{6cm}|p{3cm}|p{1cm}|p{2.5cm}|}
    \hline
    \textbf{Project} & \textbf{Description} & \textbf{Category} & \textbf{Price} & \textbf{Coordinates}\\
    \hline
    Computers for the community learning center & Funding 20 laptops including mice and keyboards, giving students a place to study & Culture and community &  27K & (55,55)\\
    \hline
    Laundry Access in Public Schools & Renovate a space in a Cambridge Public School and install washers and dryers for students who do not have easy access to laundry services at home, to use for their clothing and necessities & Culture and community & 50K & (25,45)\\
    \hline
    Real-Time Bus Arrival Monitors in bus stations & Real-time bus arrival monitors bus stops will inform travelers when the next bus will arrive, so they can adjust their plans if needed &  Streets, Sidewalks and Transit &  24K & (65,15) (35,85) (75,75)\\
    \hline
    Sheltered Bike Parking at the Main Library & The Main Library needs more bicycle parking. A glass pavilion, protecting bikes from the weather, landscaped with paths and trees, will be an attractive and functional addition to the library grounds & Streets, Sidewalks and Transit & 90K & (45,25)\\
    \hline
    24H public toilet & 24-hour access public toilet near Central Square & Environment, public health and safety & 320K & (70,35)\\
    \hline
    The Sustainable Energy Pilot & Install energy conversion devices on gym equipment and a rapid electric vehicle charging station & Environment, public health and safety & 90K & (10,25)\\
    \hline
    Dog Park & Building a dog park & Facilities, parks and recreation & 250K & (55,85)\\
    \hline
    Let’s Rest: Picnic Tables and Benches for Our Parks & Benches and picnic tables bring our community together. Installing new benches and picnic tables in up to 10 of our park will allow people of all ages and abilities to enjoy them for resting, talking, reading, people watching and being outdoors & Facilities, parks and recreation & 120K & (75,25) (90,55) (95,25)\\
    \hline
    Installing Lights at the school Basketball Court & Install lighting to extend safe playing hours for basketball courts. Increases safety for community members while expanding healthy alternatives for youth and access to public space & Education & 250K & (25,55)\\
    \hline
    Security Cameras & Install security cameras in public schools & Education & 105K & (35,10) (50,45) (40,60)\\
        
     \hline
  \caption{First small election (\textsc{Small-A})}\label{tab:first_elc}.
\end{longtable}

\begin{longtable}[ht!]{|p{2cm}|p{6cm}|p{3cm}|p{1cm}|p{2.5cm}|}
    \hline
    \textbf{Project} & \textbf{Description} & \textbf{Category} & \textbf{Price} & \textbf{Coordinates}\\
    \hline
    Interactive Technology for the Main Library & This project will fund an iPad lending kiosk and 16 iPads, as well as a permanent interactive screen in the Children’s Room of the Main Library & Culture and community & 60K & (5,5)\\
    \hline
    Laundry Access in Public Schools & Renovate a space in a Cambridge Public School and install washers and dryers for students who do not have easy access to laundry services at home, to use for their clothing and necessities & Culture and community & 50K & (90,75)\\
    \hline
    We Need More Crosswalks & To enhance pedestrian safety, this project will add a minimum of five new crosswalks to major streets & Streets, Sidewalks and Transit &  40K & (55,35) (35,25) (35,55)\\
    \hline
    Hump Installationer for pedestrians & Speed humps create a safer environment by helping slow traffic on streets that students and families cross frequently. When a car hits a pedestrian at a high rate of speed, the collision is more likely to result in a pedestrian fatality. Speed humps slow vehicles and give drivers increased response time and distance for stopping. This makes streets safer for pedestrians & Streets, Sidewalks and Transit & 66K & (75,65)\\
    \hline
    Nursing Pod for Mothers and Infants & Provide an attractive private space where working mothers and community members can breastfeed or pump during the work day & Environment, public health and safety & 20K & (55,85) (50,3) (20,90)\\
    \hline
    Soak Up the Solar Power & Free, clean, renewable energy! Let’s add solar panels to the Youth Center to reduce greenhouse gas emissions and save money on energy & Environment, public health and safety & 250K & (45,25)\\
    \hline
    Building an amphitheater in the public park & Build an amphitheater in the public park for outdoor performances, music, stories, and other cultural events that the whole community can enjoy & Facilities, parks and recreation & 350K & (70,20)\\
    \hline
    Remodel the Kitchen at the Youth Center & The kitchen area in the Youth Center is in dire need of renovating. Replace the stove, dishwasher, cabinets, and countertops in the Youth Center kitchen & Facilities, parks and recreation & 200K & (90,40)\\
    \hline
    Laptops for 10 Public Schools & Purchasing laptop carts for ten public schools & Education & 350K & (50,50)\\
    \hline
    Security Cameras & Install security cameras in public schools & Education & 105K & (20,50)\\
     \hline
  \caption{Second small election (\textsc{Small-B})}\label{tab:second_elc}.
\end{longtable}

\begin{longtable}[ht!]{|p{2cm}|p{6cm}|p{3cm}|p{1cm}|p{2.5cm}|}
    \hline
    \textbf{Project} & \textbf{Description} & \textbf{Category} & \textbf{Price} & \textbf{Coordinates}\\
    \hline
    Little (book exchange) libraries &  Installing 13 little free libraries across town & Culture and community & 13K & (1,20) (27,92) (85,35)\\
    \hline
    Computers for the community learning center & Funding 20 laptops including mice and keyboards, giving students a place to study & Culture and community &  27K & (35,60)\\
    \hline
    Digital Sign at City Hall in Multiple Languages & Digital sign that will scroll announcements in multiple languages and welcome people to town & Culture and community & 75K & (65,25)\\
    \hline
    Meeting Room Upgrade for libraries & Upgrades will allow for the latest in technology be available for public use in the meeting room & Culture and community &  250K & (5,5)\\
    \hline
    Separate Bike Lanes from Traffic & Improve safety for drivers and bikers by moving bike lanes to be between street parking spots and the sidewalk, reducing car-bike interactions and potential collisions & Streets, Sidewalks and Transit & 50K & (35,45) (55,35) (90,10)\\
    \hline
    Real-Time Bus Arrival Monitors in bus stations & Real-time bus arrival monitors bus stops will inform travelers when the next bus will arrive, so they can adjust their plans if needed &  Streets, Sidewalks and Transit &  24K & (15,65) (55,15) (85,85)\\
    \hline
    Benches for a Walkable City & Install 12 benches across town so that people of all ages and abilities can enjoy benches for resting, talking, tinkering with electronic devices, people watching, and being outdoors & Streets, Sidewalks and Transit & 25K & (75,15) (45,25) (15,45)\\
    \hline
    Urban Bicycle Wash Stations & Bicycle owners need to clean and care for their bikes, ideally monthly. But for apartment dwellers, this is really hard! These centrally located bicycle wash stations would allow bicycle owners to wash off dirt, grime, and salt from their bikes & Streets, Sidewalks and Transit & 20K & (35,35) (75,25) (85,75)\\
    \hline
    Planting trees in the city & Street trees cool the city, absorb pollution, and make our neighborhoods more livable! planting 100 new trees and building tree wells in the areas that need them most & Environment, public health and safety & 119.4K & (35,15) (65,20) (85,15)\\
    \hline
    24H public toilet & 24-hour access public toilet near Central Square & Environment, public health and safety & 320K & (10,10)\\
    \hline
    5 Water Bottle Refill Stations & At a water bottle refill station you get a healthy drink for free & Environment, public health and safety & 40K & (15,75) (15,25) (75,65)\\
    \hline
    Fire Hydrant Markers & Install high-visibility markers on fire hydrants around town to increase safety for all residents and reduce response time of the fire department by improving ease of locating hydrants in emergencies, at night, and in the snow & Environment, public health and safety & 8K & (35,75) (55,25) (92,92)\\
    \hline
    Outdoor Fitness Equipment in the public park & Install outdoor body-weight fitness equipment for stretching, strength building, and plyometric exercises & Facilities, parks and recreation & 65K & (90,50)\\
    \hline
    Volleyball Court in the public park & Creating an outdoor volleyball court would be an exciting addition to the city. The court would have sand and a sturdy net for three-season usage & Facilities, parks and recreation & 61K & (55,75)\\
    \hline
    Inclusive Playground for All Kids & This Universal Design playground would include equipment that is designed to be usable by everyone without special adaptations or retrofitting & Facilities, parks and recreation & 305K & (55,60)\\
    \hline
    Protect the Health and Safety of our Firefighters & This proposal will purchase and install six gear drying units to shorten wait time for clean gear (\$55,000), and eleven sets of wireless headsets to protect hearing and improve communication (\$55,000). Let’s protect those who protect us & Facilities, parks and recreation & 110K & (50,50)\\
    \hline
    New Chairs for Public Schools & New Chairs for Public Schools & Education & 190K & (90,75)\\
    \hline
    Invention and Production of Music & Install music studios and equipment at the Youth Centers to inspire creativity, enable pre-teens and teens to express their skills and passions, and provide youth with another recreational outlet & Education & 150K & (35,85)\\
    \hline
    Upgraded Water Fountains for Public Schools & Project would install 35 new water bottle refilling fountains in public schools & Education & 200K & (70,90)\\
    \hline
    New Playground for public school & Playground should include a jungle gym, and other equipment for kids to play different games & Education & 200K & (50,1)\\

     \hline
  \caption{First big election (\textsc{Large-A})}\label{tab:third_elc}.
\end{longtable}

\begin{longtable}[ht!]{|p{2cm}|p{6cm}|p{3cm}|p{1cm}|p{2.5cm}|}
    \hline
    \textbf{Project} & \textbf{Description} & \textbf{Category} & \textbf{Price} & \textbf{Coordinates}\\
    \hline
    Little (book exchange) libraries &  Installing 13 little free libraries across town & Culture and community & 13K & (85,60) (85,35) (28,28)\\
    \hline
    Computers for the community learning center & Funding 20 laptops including mice and keyboards, giving students a place to study & Culture and community &  27K & (45,25)\\
    \hline
    Meeting Room Upgrade for libraries & Upgrades will allow for the latest in technology be available for public use in the meeting room & Culture and community &  250K & (15,10)\\
    \hline
    Accessible Entrance for the Library & Add automatic sliding doors, fix driveway and, if needed, remove steps to benefit seniors and people with disabilities & Culture and community & 250K & (5,5)\\
    \hline
    Bike repair stations & Install 8 bike repair stations with tools and bike pumps across the city for cyclists to quickly, easily, and freely fix routine bike problems & Streets, Sidewalks and Transit & 12K & (35,80) (75,25) (40,40)\\
    \hline
    Separate Bike Lanes from Traffic & Improve safety for drivers and bikers by moving bike lanes to be between street parking spots and the sidewalk, reducing car-bike interactions and potential collisions & Streets, Sidewalks and Transit & 50K & (20,33) (55,35) (10,50)\\
    \hline
    Sheltered Bike Parking at the Main Library & The Main Library needs more bicycle parking. A glass pavilion, protecting bikes from the weather, landscaped with paths and trees, will be an attractive and functional addition to the library grounds & Streets, Sidewalks and Transit & 90K & (90,90)\\
    \hline
    What’s Your Speed? & Remind drivers to slow down by deploying live speed displays on busiest streets &  Streets, Sidewalks and Transit & 75K & (35,5) (35,45) (32,65)\\
    \hline
    24H public toilet & 24-hour access public toilet near Central Square & Environment, public health and safety & 320K & (45,45)\\
    \hline
    Flashing Crosswalks for Safer Streets & This project would fund rapid flashing beacons at 10 high pedestrian risk crosswalks. These beacons increase the visibility of pedestrians, especially at night. They can alert drivers to crossing pedestrians, thereby preventing crashes & Environment, public health and safety & 176K & (35,35) (62,90) (15,75)\\
    \hline
    Soak Up the Solar Power & Free, clean, renewable energy! Let’s add solar panels to the Youth Center to reduce greenhouse gas emissions and save money on energy & Environment, public health and safety & 250K & (45,15)\\
    \hline
    Fire Hydrant Markers & Install high-visibility markers on fire hydrants around town to increase safety for all residents and reduce response time of the fire department by improving ease of locating hydrants in emergencies, at night, and in the snow & Environment, public health and safety & 8K & (35,75) (85,1) (15,45)\\
    \hline
    Free Wifi in 6 Outdoor Public Spaces & Install special outdoor wifi access points to offer free public wifi in the public space & Facilities, parks and recreation & 42K & (25,90) (1,25) (80,80)\\
    \hline
    Inclusive Playground for All Kids & This Universal Design playground would include equipment that is designed to be usable by everyone without special adaptations or retrofitting & Facilities, parks and recreation & 305K & (55,75)\\
    \hline
    Shade and Wet Weather Canopies for Playgrounds & Installing canopies over playgrounds that do not have protection from the elements will reduce weather-related safety concerns and increase playground availability and use & Facilities, parks and recreation & 146K & (75,65) (45,90) (25,75)\\
    \hline
    Let’s Rest: Picnic Tables and Benches for Our Parks & Benches and picnic tables bring our community together. Installing new benches and picnic tables in up to 10 of our park will allow people of all ages and abilities to enjoy them for resting, talking, reading, people watching and being outdoors & Facilities, parks and recreation & 120K & (92,55) (60,21) (72,40)\\
    \hline
    New Chairs for Public Schools & New Chairs for Public Schools & Education & 190K & (90,75)\\
    \hline
    Installing Lights at the school Basketball Court & Install lighting to extend safe playing hours for basketball courts. Increases safety for community members while expanding healthy alternatives for youth and access to public space & Education & 250K & (85,10)\\
    \hline
    Upgraded Water Fountains for Public Schools & Project would install 35 new water bottle refilling fountains in public schools & Education & 200K & (60,5)\\
    \hline
    Security Cameras & Install security cameras in public schools & Education & 105K & (10,60)\\

     \hline
  \caption{Second big election (\textsc{Large-B})}\label{tab:fourth_elc}.
\end{longtable}


\section{Survey Questions}
Each participant was required to rate on a scale of 1 to 5 the following:

\begin{itemize}
    \item "How easy did you find the voting task?"
    \item "How much did you like the user interface?"
    \item "How well did the input format capture your preferences?"
    \item "How much did the map affect your decisions?"
    \item "How much did the project categories affect your decisions?"
    \item "How easy was it to access the map?"
\end{itemize}

\section{Consistency Questions}
Each participant was given three consistency questions with four choices:
\begin{itemize}
    \item "What was your mission?" - the participant was shown four definitions of the input formats (as defined in the tutorial) and was required to choose the one that relate to his task.
    \item "Which of the following projects did not appear in the list?" - the participant was shown three projects that were part of the election and a fourth option "Improving parking at the airport". 
    \item A question about the projects the participant voted for:
    \begin{itemize}
        \item Approval based: "Which of the following projects did you select?" - the participant was shown three projects that did not approve, and one that they did. 
        \item Points based: "To which of the following projects did you allocate the most points?" - the participant was shown the project that was given the highest points and three other projects.
        \item Rank based: "Which of the following projects did you rank the highest?" - the participant was shown the project that was ranked at the top and three other projects.
    \end{itemize}
\end{itemize}

\section{Demographic Statistics}
\begin{figure}[!h]
\begin{center}
\includegraphics[width=13cm,height=3cm]{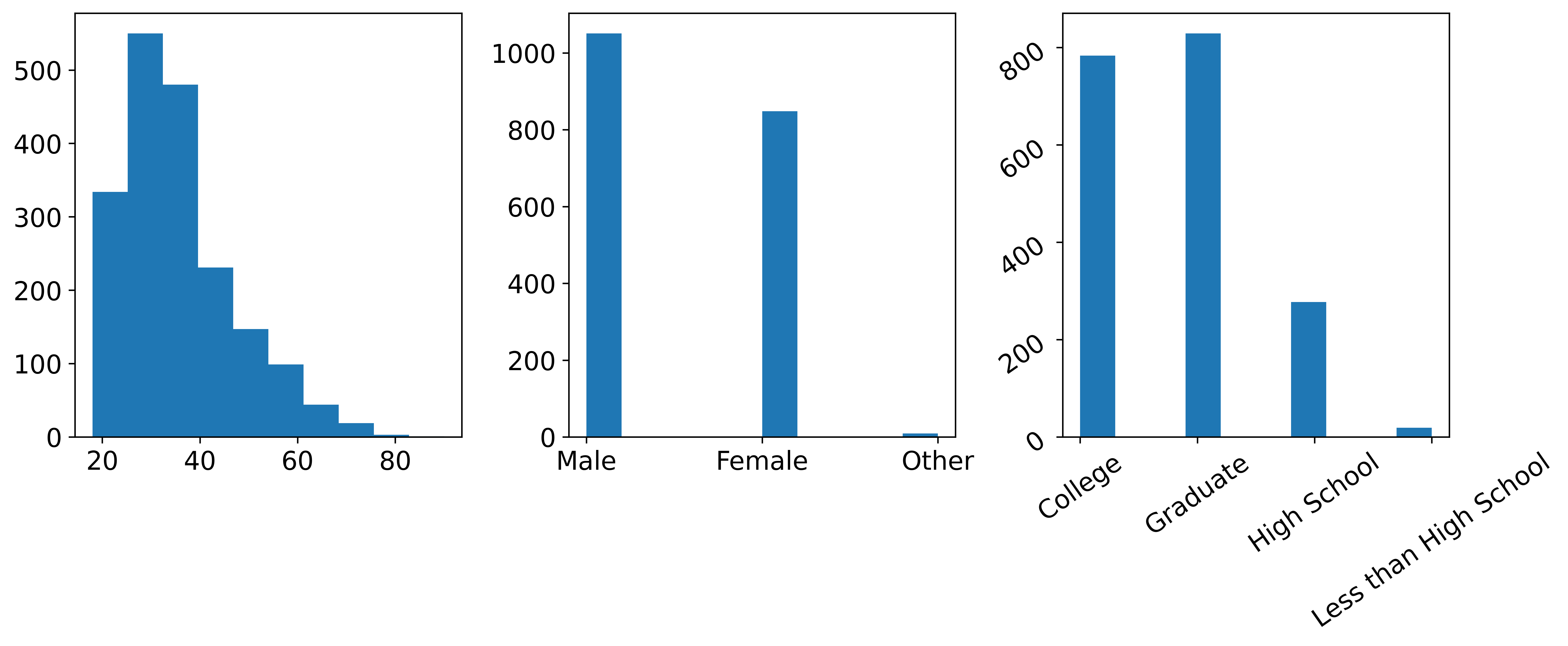}
\caption{Demographic age (left), gender (middle) and education (right) distribution.
}\label{fig:distribution}
\end{center}
\end{figure}

As part of the experiment, each participant reported their age, gender and education. About 45\% were female with ages ranging from 20-80 (mode ~30, mean ~35). Roughly 85\% graduated from (or currently in) college. More detailed distribution can be seen in Figure~\ref{fig:distribution}.

\section{Details omitted from Comparing input formats}
\subsection{Response time}
\begin{figure}[!h]
\begin{center}
\includegraphics[width=7.5cm]{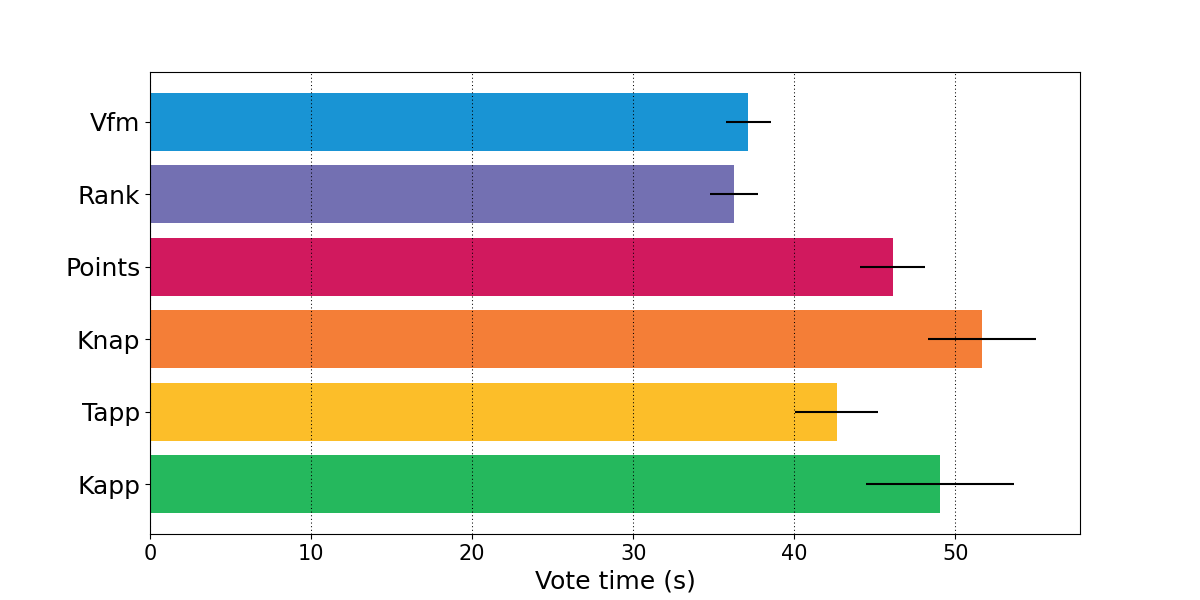}
\caption{How long it took the participants to do the consistency quiz (in seconds) on average.
}\label{fig:consistency_time}
\end{center}
\end{figure}

Figure~\ref{fig:consistency_time} shows how long the consistency stage took. 

\subsection{Consistency}
\begin{figure*}[ht!]
     \centering
          \begin{subfigure}[b]{0.45\textwidth}
         \centering
       \includegraphics[width=6cm]{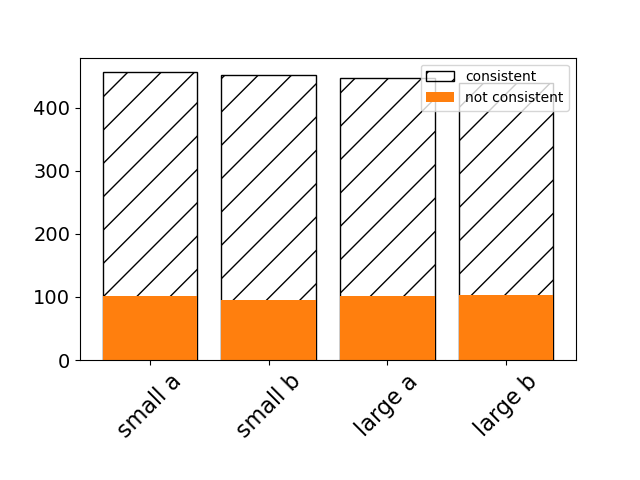}
\caption{The total number of  participants who vote in each election, as well as the number that fail the consistency test.  Numbers are aggregated across formats.
}\label{fig:consistency_elections}
     \end{subfigure}\hfill
     \begin{subfigure}[b]{0.45\textwidth}
         \centering
         \includegraphics[width=6cm]{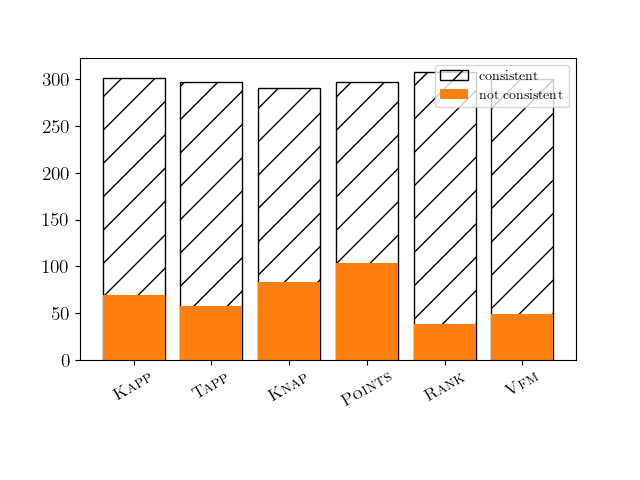}
\caption{The total number of  participants who vote in each format, as well as the number that fail the consistency test.  Numbers are aggregated across elections.
}\label{fig:consistency}
     \end{subfigure}
        \caption{Input formats and elections consistency.}
        \label{fig:all_consistency}
\end{figure*}

Figures~\ref{fig:consistency_elections},\ref{fig:consistency} how many participants were consistent across different formats and across different project sets.


\section{Entropy}\label{app:entropy}
\begin{figure*}[!h]
     \centering
          \begin{subfigure}[b]{0.45\textwidth}
         \centering
        \includegraphics[width=6cm]{experiment/entropy_small_a.png}
\caption{Greedy (full lines) and \mes (dotted lines) entropy  for each input format (sample format get same color) given different sample size in project set \textsc{Small-A} }\label{fig:entropy_sa}
     \end{subfigure}\hfill
     \begin{subfigure}[b]{0.45\textwidth}
         \centering
         \includegraphics[width=6cm]{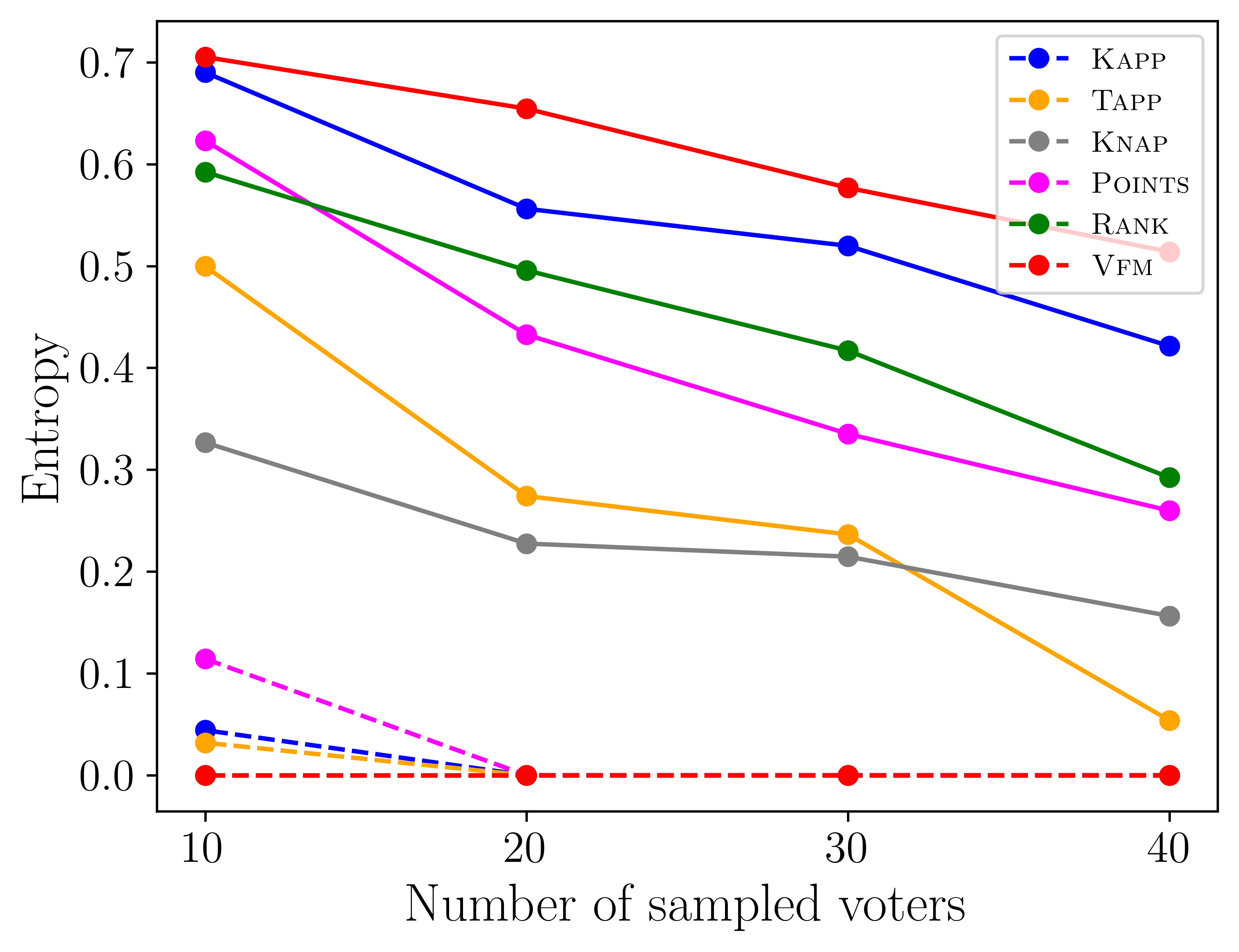}
\caption{Greedy (full lines) and \mes (dotted lines) entropy  for each input format (sample format get same color) given different sample size in project set \textsc{Small-B}}\label{fig:entropy_sb}
     \end{subfigure}
     \hfill
     \begin{subfigure}[b]{0.45\textwidth}
         \centering
         \includegraphics[width=6cm]{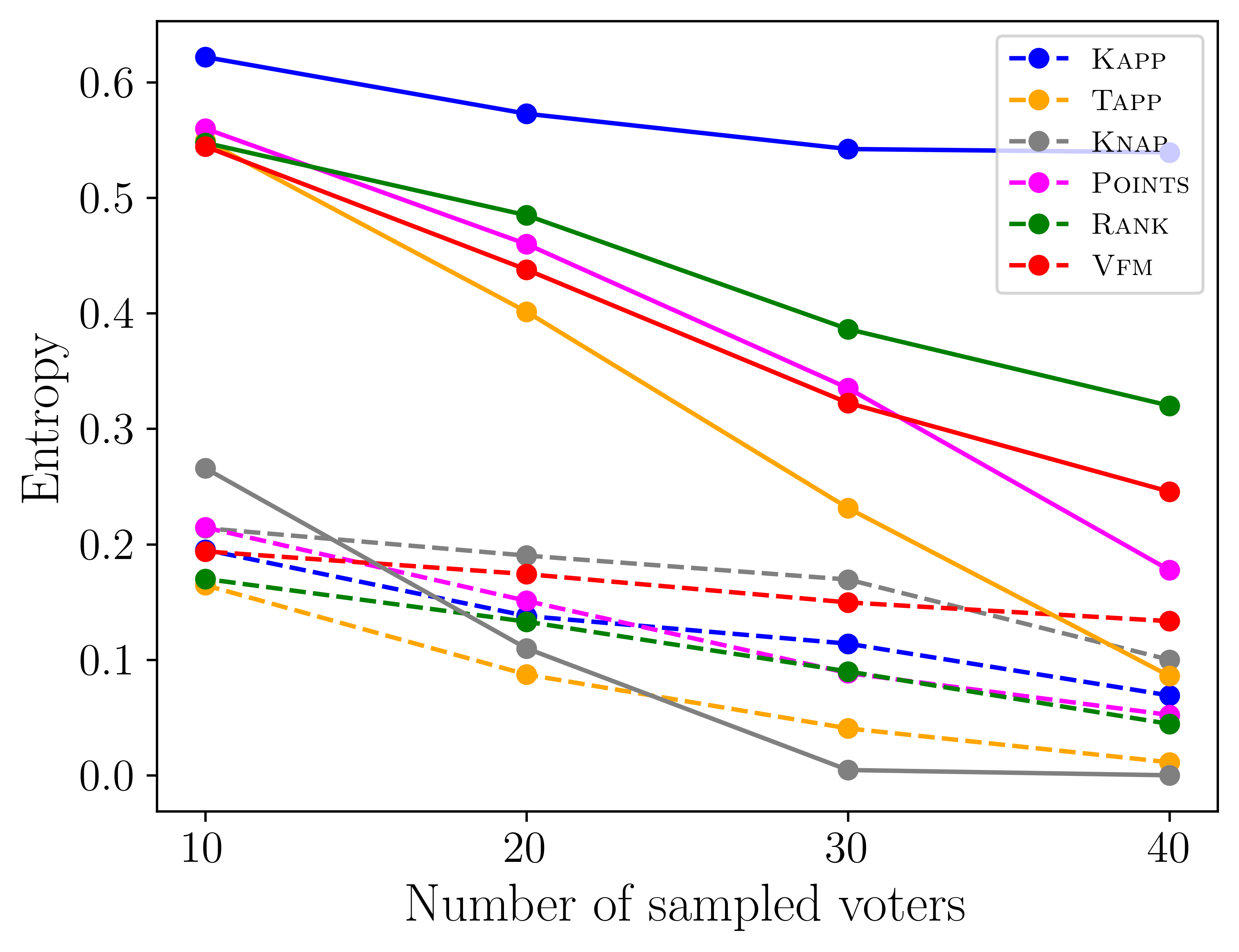}
\caption{Greedy (full lines) and \mes (dotted lines) entropy  for each input format (sample format get same color) given different sample size in project set \textsc{Large-A} }\label{fig:entropy_la}
     \end{subfigure}
     \hfill
     \begin{subfigure}[b]{0.45\textwidth}
         \centering
         \includegraphics[width=6cm]{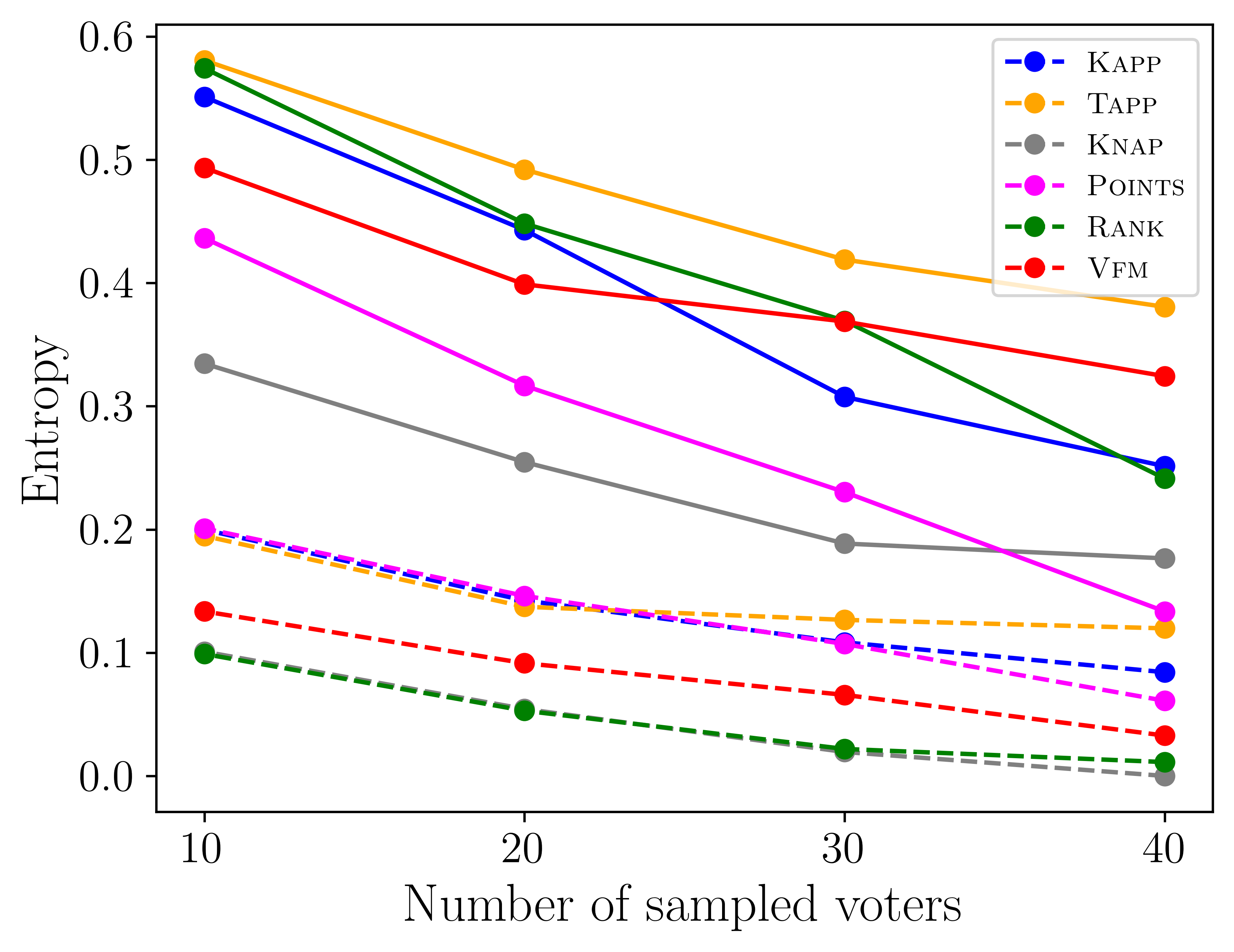}
\caption{Greedy (full lines) and \mes (dotted lines) entropy  for each input format (sample format get same color) given different sample size in project set \textsc{Large-B} }\label{fig:entropy_lb}
     \end{subfigure}

        \caption{Entropy for each of the elections.}
        \label{fig:app_entropy}
\end{figure*}

Figures~\ref{fig:entropy_sa}-\ref{fig:entropy_lb}  shows the entropy for all of the project sets, all input formats when using greedy and \mes{}, given different sample size.

\section{Welfare from all formats voters}\label{app:aggregation}
\begin{figure*}[ht]
     \centering
          \begin{subfigure}[b]{0.45\textwidth}
         \centering
       \includegraphics[width=6cm]{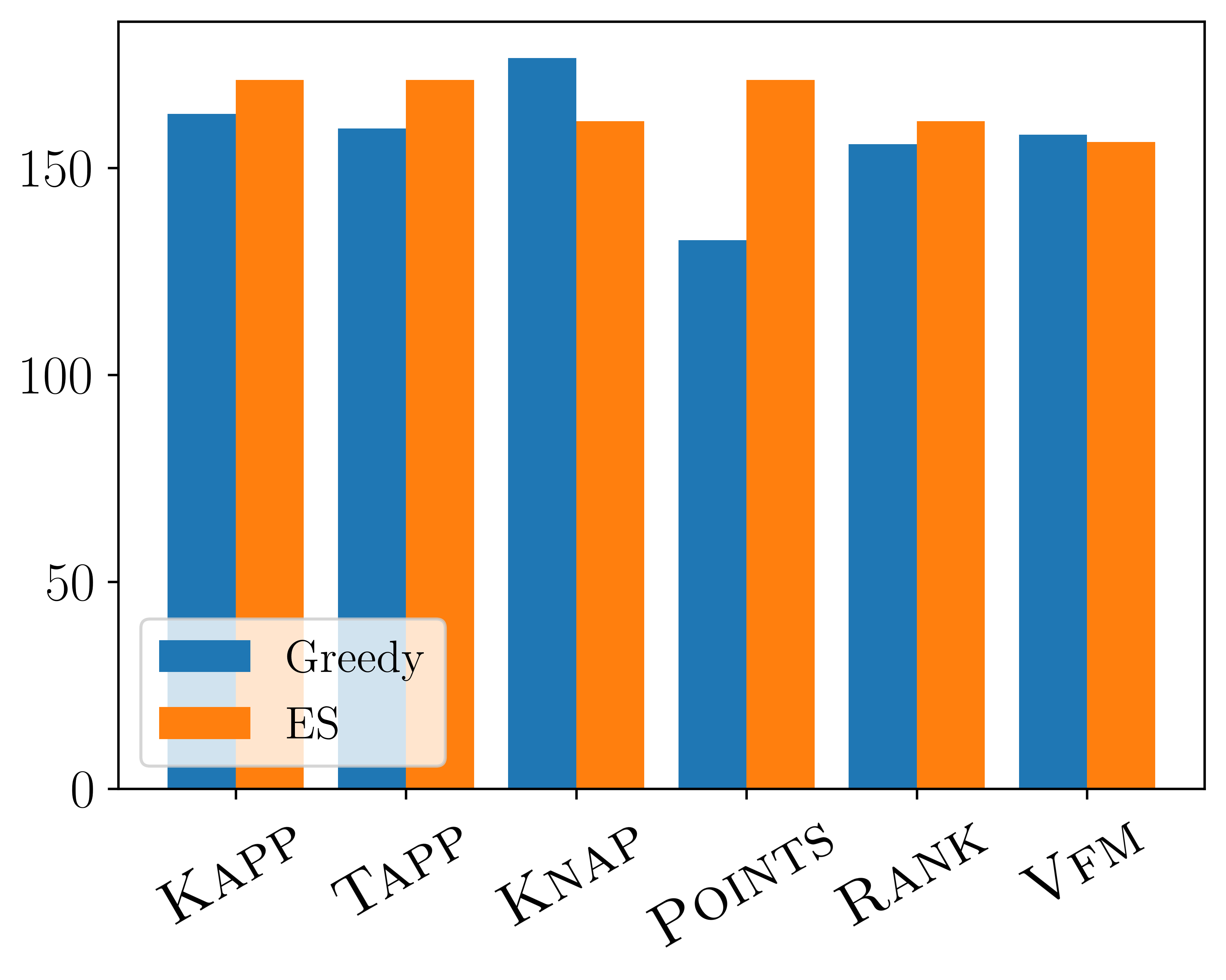}
\caption{Average social welfare (averaged across elections) using \kapp{}-voters to compute welfare.
}\label{fig:kapp_welfare}
     \end{subfigure}\hfill
     \begin{subfigure}[b]{0.45\textwidth}
         \centering
         \includegraphics[width=6cm]{experiment/Knapsack_welfare.png}
\caption{Average social welfare (averaged across elections) using \knap{}-voters to compute welfare.
}\label{fig:knap_welfare}
     \end{subfigure}
     \hfill
     \begin{subfigure}[b]{0.45\textwidth}
         \centering
         \includegraphics[width=6cm]{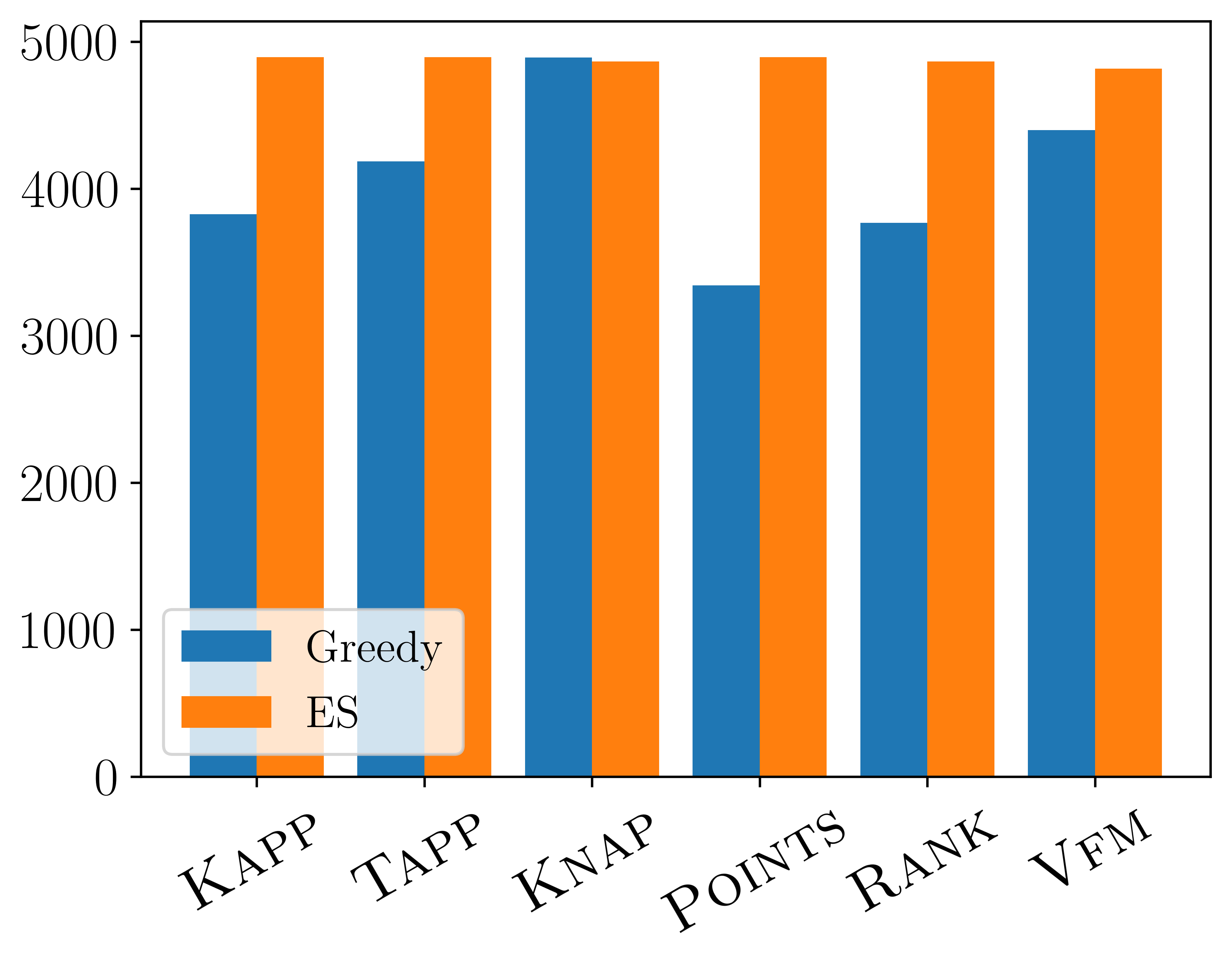}
\caption{Average social welfare (averaged across elections) using \rank{}-voters to compute welfare.
}\label{fig:rank_welfare}
     \end{subfigure}
     \hfill
     \begin{subfigure}[b]{0.45\textwidth}
         \centering
\includegraphics[width=6cm]{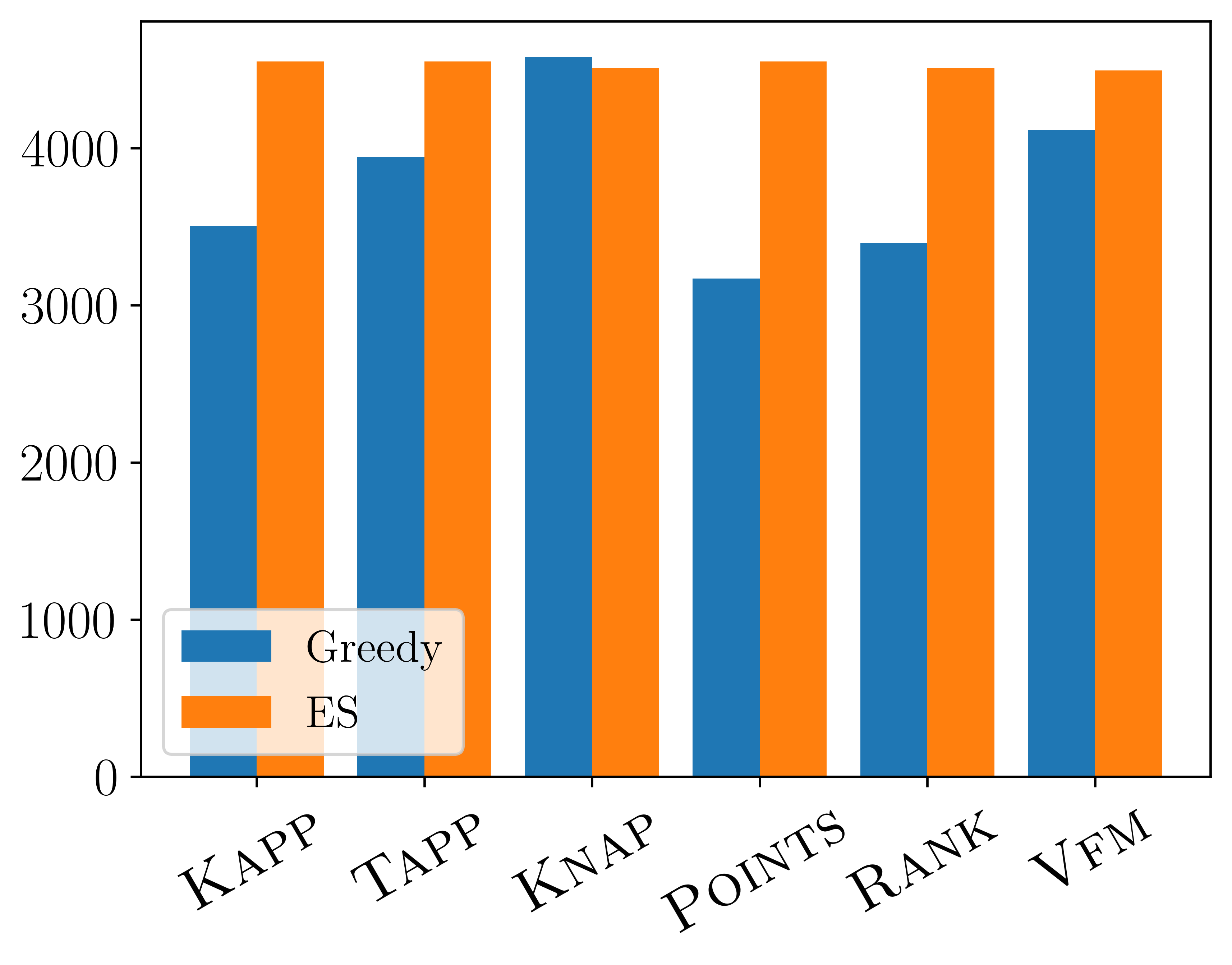}
\caption{Average social welfare (averaged across elections) using \vfm{}-voters to compute welfare.
}\label{fig:vfm_welfare}
     \end{subfigure}
     \begin{subfigure}[b]{0.45\textwidth}
         \centering
\includegraphics[width=6cm]{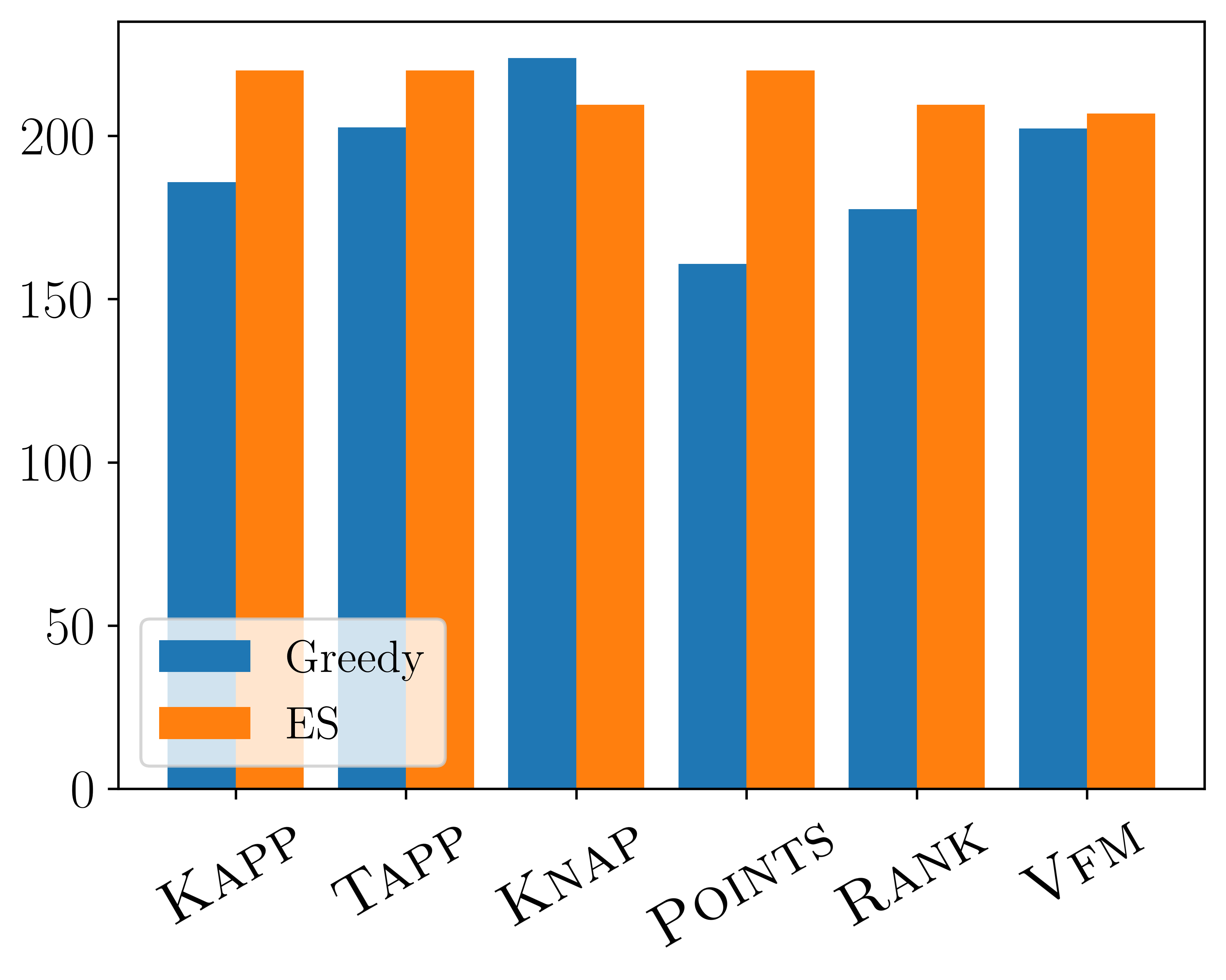}
\caption{Average social welfare (averaged across elections) using \tapp{}-voters to compute welfare.
}\label{fig:tapp_welfare}
     \end{subfigure}
     \hfill
     \begin{subfigure}[b]{0.45\textwidth}
         \centering
\includegraphics[width=6cm]{experiment/Utilities_welfare.png}
\caption{Average social welfare (averaged across elections) using \points{}-voters to compute welfare.
}\label{fig:util_welfare}
     \end{subfigure}

        \caption{Welfare.}
        \label{fig:entropy:app}
\end{figure*}

Figures~\ref{fig:kapp_welfare}-\ref{fig:util_welfare} shows the welfare (averaged across elections), when using each of the input formats voters to calculate the welfare.







\end{document}